\documentclass[12pt]{article}
\usepackage{epsfig}
\graphicspath{{./}{./pdf}{./eps}{./figures/}{./figures/eps/}{./figures/pdf/}}
\usepackage{color}
\definecolor{refblue}{rgb}{0.,.17,.45}
    \def\nuc#1#2{\relax\ifmmode{}^{#1}{\protect\text{#2}}\else${}^{#1}$#2\fi}
    \newcommand{\be}{\begin{equation}}
    \newcommand{\ee}{\end{equation}}
    \newcommand{\bea}{\begin{eqnarray}}
    \newcommand{\eea}{\end{eqnarray}}

\begin{document}

\title{\vspace{1cm} Fusion and Direct Reactions of Halo Nuclei\\
at Energies around the Coulomb Barrier}

\author{N.\ Keeley$^1$, R.\ Raabe$^2$, N.\ Alamanos$^1$, J. L.\
Sida$^3$\\ \\
$^1$CEA DSM/DAPNIA/SPhN, F-91191 Gif-sur-Yvette, France\\
$^2$Instituut voor Kern- en Stralingsfysica, K.U.Leuven,\\
Celestijnenlaan 200 D, B-3001 Leuven, Belgium\\
$^3$CEA DIF/DPTA/Service de Physique Nucl{\'e}aire,\\ F-91680
Bruy{\`e}res-le-Ch{\^a}tel, France\\}

\maketitle

\begin{abstract}
The present understanding of reaction processes involving light
unstable nuclei at energies around the Coulomb barrier is reviewed.
The effect of coupling to direct reaction channels on elastic
scattering and fusion is investigated, with the focus on halo
nuclei, for which such effects are expected to be most important.
With the aim of resolving possible ambiguities in the terminology a
short list of definitions for the relevant processes and quantities
is proposed. This is followed by a review of the experimental and
theoretical tools and information presently available. The effect
of breakup couplings on elastic scattering and of transfer
couplings on fusion is investigated with a series of model
calculations within the coupled-channels framework. The
experimental data on fusion are then compared to ``bare''
no-coupling one-dimensional barrier penetration model calculations
employing reasonably realistic double-folded potentials. On the
basis of these model calculations and comparisons with experimental
data, conclusions are drawn from the observation of recurring
features. The total fusion cross sections for halo nuclei show a
suppression with respect to the ``bare'' calculations at energies
just above the barrier that is probably due to single neutron
transfer reactions. The data for total fusion are also consistent
with a possible sub-barrier enhancement; however, this observation
is not conclusive and other couplings besides the single-neutron
channels would be needed in order to explain any actual
enhancement. We find that a characteristic feature of halo nuclei
is the dominance of direct reactions over fusion at near and
sub-barrier energies; the main part of the cross section is related
to neutron transfers, while calculations indicate only a modest
contribution from the breakup process.
\end{abstract}

\eject

\tableofcontents

%\newpage

\section{Introduction
\label{sec:intro}}

Collisions between two heavy nuclei can lead to a variety of
processes. In a semiclassical picture, it is customary to use the
impact parameter, or relative angular momentum, to distinguish
between \emph{compound nucleus} reactions (fusion reactions) and
\emph{direct} reactions; the latter take place for values of the
impact parameter corresponding to grazing trajectories, the former
occur for more central collisions. In between, \emph{deep
inelastic} reactions show intermediate behaviour. The relative
importance of the different mechanisms depends on a number of
factors; among these, a very important r\^ole is played by the
kinetic energy and its value with respect to the amount of Coulomb
repulsion between the nuclei for a given trajectory (the ``Coulomb
barrier''). At relatively high energies with respect to the barrier
it is possible to use geometrical models of the reaction process to
provide independent descriptions of individual mechanisms: for
example, Glauber models \cite{Gla55} for direct reactions, the
one-dimensional barrier penetration model for fusion \cite{Won73}.

When the kinetic energy is small compared to the Coulomb barrier
height this independence no longer holds: the behaviour of a
particular process can no longer be considered separately from the
others. In scattering theory this is expressed by the concept of
\emph{coupling} of the different \emph{reaction channels}.
Formally, the total wave function of the scattering problem
contains the entrance channel and all possible exit channels; the
Hamiltonian connects these states by means of potential terms, for
example potentials that can create an excitation. For small kinetic
energies, the contributions of these terms become significant in
the determination of the wave function of each channel. The effect
of the couplings is well-established, and visible on both the
elastic scattering \cite{Sat90} and fusion cross sections
\cite{Bec85,Ste86}. In this review we investigate the effect of
such couplings when one of the participants in the reaction is a
light, unstable nucleus.

As recognized since the very first measurements in the mid-eighties
\cite{Tan85,Tan85a}, light exotic nuclei have properties which
influence the reaction mechanism. Even models for high energy
collisions must take into account the cluster structure of these
nuclei \cite{AlK96a} and their extended mass distributions. At
energies around the Coulomb barrier the couplings between the
various reaction channels are expected to be particularly
important. This derives from the fact that some direct channels may
be enhanced due to the characteristics of these systems: for
example, selected transfer channels may be favoured by large
positive $Q$-values and the cluster structure of the exotic
nucleus; a low breakup threshold may cause a large cross section
for the breakup channel.

To test these conjectures, more than ten years were necessary in
order to develop suitable beams of radioactive nuclei in the energy
range of interest and the experimental techniques to deal with the
very low currents of such beams. The first measurements of the
fusion cross section with light unstable projectiles were made in
the second half of the nineties. Great interest in the fusion
reaction mechanism was driven by the prediction of a possibly
exceptional enhancement at sub-barrier energies when using a halo
nucleus as projectile. The large spatial extension of the neutron
matter density would cause the strong interaction to set in at
larger distances, reducing the effect of the Coulomb repulsion. In
addition, for these nuclei special couplings would appear: the
presence of a large part of the dipole strength at low excitation
energy, possibly concentrated into a resonance, would favour the
corresponding excitation channel. However, as the breakup threshold
in halo nuclei is very low the effect of the breakup channel
becomes crucial.

An intense debate arose about the possible effect of breakup,
mainly due to the way the mechanism was described. A low breakup
threshold implies that the Coulomb or nuclear interactions can
easily split the exotic nucleus into its cluster components. In one
picture \cite{Hus92}, this would enhance the breakup probability
and, in turn, hinder the fusion of the whole exotic projectile with
the target nucleus, due to a decrease of the number of nuclei
available for fusion. This point of view assumes a two-step
mechanism, a position which has been strongly objected to by other
authors on grounds that are well described for example in
\cite{Das94}. They argue that a decrease in the observed flux
reflected from the entrance channel (i.e. a decrease of the elastic
scattering due to breakup) does not imply a decrease of the flux
available for fusion: the latter can in fact even increase, since
the transmitted flux is determined by a different region of the
potential (more inside the barrier) with respect to the reflected
flux. The problem can be properly treated only in a full
coupled-channels picture. In this picture, coupling to other
reaction channels at sub-barrier energies would lead to an
enhancement of the transmitted flux and thus of the fusion cross
section. This is the equivalent, at the nuclear level, of the
enhancement of the tunnelling probability due to the presence of
additional degrees of freedom in a given system, as described for
example in \cite{Cal81}.

The quantum-mechanical scattering problem should indeed be treated
in its totality, and the coupled channels method is directly
derived from the Schr\"odinger equation related to the problem.
However, the inclusion of couplings is not always straightforward,
and this has been the reason behind the formulation of alternative
descriptions. The inclusion of the breakup channel represents a
particular challenge. Breakup is essentially an excitation of the
nucleus to an energy higher than the corresponding threshold; if
the calculation treats it in the same way as an inelastic
excitation to a bound state, it will indeed lead to an enhancement
of the tunnelling probability. However, the excitation is into the
continuum, in many cases not even to a resonance and thus with no
well defined values for the energy or spin of the final state. The
effect of such a coupling cannot be predicted \emph{a priori}.

A method to calculate the fusion cross section including the effect
of breakup in a coupled-channels picture has not yet been fully
developed. In an attempt to improve our understanding of the
problem, we shall investigate the effect of breakup as a function
of threshold energy and the target nucleus on the elastic
scattering, a problem for which a realistic treatment is available
(the Continuum-Discretised Coupled Channels method, CDCC).

In addition to breakup, we shall devote our attention to transfer
channels and their effects. These are also expected to be large for
light exotic nuclei, and indeed have recently been measured to be
so. Transfer channels can be taken into account when calculating
fusion in a coupled-channels framework, thus a detailed
investigation is in this case possible.

Experimental investigation of reactions involving light exotic
nuclei has been largely limited to those systems having a halo
structure, i.e.\ essentially \nuc{6}{He} and \nuc{11}{Be} as the
other halo nuclei (e.g.\ \nuc{11}{Li} and \nuc{14}{Be}) are not yet
available as beams with sufficient intensity. The effects due to
breakup and transfer channels should also be largest for halo
nuclei, due to their very low breakup thresholds and cluster
structures. Since we wish to base our conclusions on the available
experimental results we shall therefore be mainly dealing with halo
nuclei. We choose to compare the measured fusion cross sections
with calculations made with a bare potential obtained by folding
the densities of the nuclei, and formulate our observations in view
of this comparison and the results obtained from the
coupled-channels calculations. Also, we limit ourselves to the very
light systems up to $Z=4$.

The structure of the paper is as follows. In
section~\ref{sec:definitions} we give a short list of definitions
for the processes and quantities of interest, with the intention of
aiding the clarity of the exposition. Even though the scope of the
definitions is limited to this paper, we believe that this is a
first, necessary step in trying to solve some of the ambiguities in
this field. We then explain why, for our discussion, we choose to
compare the experimental data to a fusion calculation obtained
using a bare potential. In section~\ref{sec:exp} we present the
techniques used in measurements with light unstable nuclei, and
review the results for those systems for which fusion has been
measured. We then discuss the coupled-channels methods for
calculation of the fusion and direct reaction cross sections,
describing ways of including the breakup channel. We use these
methods for a series of model calculations for breakup and its
effect on elastic scattering, and transfer and its effect on
fusion. The available experimental data are consistently compared
with model calculations for fusion in section~\ref{sec:discussion}
and general deductions are drawn. Finally, our conclusions are
summarised in section~\ref{sec:conclusions}.

%\newpage

\section{Definitions
\label{sec:definitions}}

\subsection{Definition of reaction processes
\label{subsec:def_proc}}

Halo nuclei have exhibited remarkable features since the
measurement of the total interaction cross section in 1985
\cite{Tan85,Tan85a}. Characteristics such as cluster structures and
weak binding energies suggested that reaction mechanisms which were
either usually neglected in reactions involving ``normal'' nuclei
or previously unsuspected could be important. Unfortunately, the
terms chosen to designate such processes have not always been used
in a consistent way by all authors. In addition, some processes are
difficult to define unambiguously. This has led to a sometimes
confusing situation in the current literature, and certain
discrepancies in reported results may be connected with this
problem.

In this section we present a short list of definitions for use in
the current paper. The aim is to adopt a uniform vocabulary in the
review of experimental results and in the presentation of
theoretical models, in order to facilitate comparison and help the
clarity of the discussion.

\begin{itemize}

\item \textbf{Coulomb barrier} The name given to the energy
threshold that two nuclei must overcome in order to come close
enough together to fuse classically. The Coulomb barrier (sometimes
simply called the \emph{potential barrier}) has a simple
interpretation in the case of a one-dimensional potential, only
depending upon the radial separation of the interacting nuclei. For
a central collision the total potential is then given by the sum of
the repulsive Coulomb and the attractive short-range nuclear
potentials, see Fig.\ \ref{fig:potential}. The \emph{barrier
energy} is then the maximum of the combined potential,
$V_\mathrm{B}$, and its position is the \emph{barrier radius},
$R_\mathrm{B}$. For centre-of-mass collision energies lower than
$V_\mathrm{B}$ fusion can take place only by quantum-mechanical
tunnelling. If the collision is not central, the angular momentum
contributes an additional centrifugal barrier. For large values of
the angular momentum, the nuclear force is not sufficient to create
a ``pocket'' and the potential is repulsive for all distances, see
Fig.\ \ref{fig:potential}.
\begin{figure}
\epsfysize=9.0cm
\begin{center}
    \includegraphics[width=.5\textwidth]{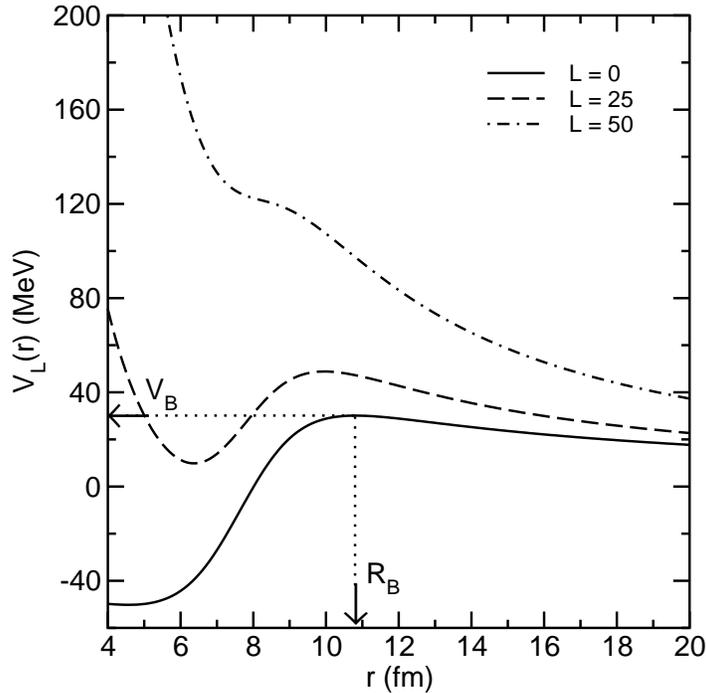}
\end{center}
\caption{The interaction potential for $^7$Li + $^{208}$Pb as a
function of radial separation between the centres of the two
nuclei for representative partial waves.\label{fig:potential}}
\end{figure}
Many results may be interpreted in terms of modifications to the
height, position and shape of the barrier, due to the presence of
strong direct reaction channels.

\item \textbf{Transfer} A rearrangement process whereby one or
more nucleons are transferred from the projectile to the target
nucleus or \emph{vice versa}. A direct process, thus peripheral,
taking place for angular momenta around the semiclassical critical
value (corresponding to grazing trajectories). The resulting
nucleus may be formed in its ground state or in bound or unbound
(resonant) excited states. If formed in an excited state the
residual nucleus may decay via the emission of $\gamma$ rays (bound
and some resonant states) or neutrons, protons or other charged
particles (resonant states), or it may fission.
\newline Transfer as intended here should be distinguished from
the mass transfer occurring in \emph{quasi-fission}
processes~\cite{Swi81}: the latter are heavy-ion dissipative
reactions, in which exchange of nucleons leads to a deformed
composite system which subsequently decays into fission-like
fragments.

\item \textbf{Breakup} A direct process where the
projectile is split into two or more fragments, due to the Coulomb
and/or nuclear interactions. A practical definition is the cross
section obtained in a coincidence measurement, i.e.\ for the
breakup process $^6$Li $\rightarrow$ $\alpha + d$ the breakup cross
section is defined as the $\alpha + d$ coincidence cross section.
This definition thus excludes events where breakup is followed by
capture of one or more of the fragments and is analogous to the
definition of inelastic excitation of a bound excited state. We
further restrict our definition to those coincidence events where
the target remains in its ground state, corresponding to what is
sometimes referred to in the literature as ``elastic breakup'', the
target nucleus remaining a spectator during the breakup process
(see e.g.\ \cite{Pam78,Uda80,Tak81}).

\item \textbf{Fusion} The process whereby two nuclei
collide and coalesce to form a \emph{compound system}. The term
fusion is used to indicate any \emph{compound nucleus process}
between two heavy ions: the fusing nucleons lose the ``memory'' of
the nucleus from which they came. Note that all the nucleons from
both systems do not necessarily take part in the fusion process.
The compound system initially has some excitation energy, and
reaches equilibrium by emitting various kinds of prompt radiation:
$\gamma$ rays, light charged particles, neutrons, or possibly by
fission. The heaviest remnant is the \emph{residual nucleus}; this
may in turn be unstable and emit further delayed radiation
($\alpha$,$\beta$,$\gamma$) that can be detected for its
identification.

\item \textbf{Complete fusion} In this process, \emph{all} nucleons
from the projectile and the target fuse into the compound system.
This quantity can be calculated by some models, but its
unambiguous measurement is, in the general case, very difficult
(see the discussion in section~\ref{subsec:identification}).

\item \textbf{Incomplete fusion} This is the fusion of a large
part of the projectile with the target nucleus. With light nuclei,
for which only few nucleons are involved, incomplete fusion and
transfer reactions remain conceptually different mechanisms but can
lead to the formation of the same final nucleus in a similar
excitation state. The angular momenta involved in the two processes
are also similar, since the (semiclassical) value of the critical
angular momentum is small. Thus, the separation between incomplete
fusion and direct nucleon transfer may not even be significant
experimentally, though attempts in this direction have been
reported \cite{Tri05}. Incomplete fusion has been described in a
two-step process picture as \textbf{fusion following breakup} where
the projectile is first broken into two or more fragments by
Coulomb and/or nuclear forces and some of the fragments penetrate
the barrier and fuse with the target.
%A special case of incomplete fusion relevant to weakly bound
%nuclei is \textbf{fusion following breakup} where the projectile
%is first broken into two or more fragments by Coulomb and/or
%nuclear forces and some of the fragments penetrate the barrier and
%fuse with the target.
Strictly speaking, events where \emph{all} fragments fuse with the
target nucleus after breakup are also possible (and experimentally
indistinguishable from complete fusion). However, they must be
considered as rather unlikely, although see \cite{Hag04}. In the
introduction we discussed briefly the two-step picture as opposed
to the coupled-channels one. In this respect, fusion following
breakup should be distinguished from \textbf{sequential transfer}
processes, which is the name given to those transfer probabilities
which are coupled to an excitation to an unbound resonant state or
the non-resonant continuum.
\newline Direct cluster transfer from the ground state of a light
weakly bound nucleus can also produce the same residual nuclei as
fusion following breakup and sequential transfer. Such processes
need to be carefully distinguished from fusion events, see e.g.\
\cite{Raa04,DiP04,Nav04} and the discussion in
section~\ref{subsec:identification}.

%Fusion following breakup is not easily distinguished from
%\textbf{sequential transfer} processes where the projectile is
%first excited to an unbound resonant state or the non-resonant
%continuum before transfer takes place.

\item \textbf{Total fusion} This is the sum of complete fusion and
incomplete fusion. The complete projectile or one or more
fragments fuse with the target. In many cases this is the simplest
process to measure, as it is an inclusive cross section.

\item \textbf{Fusion enhancement or suppression} There are two main
schools of thought in the literature as to how to define whether
fusion --- in particular sub-barrier fusion --- is  enhanced or
suppressed. The original definitions, used for stable, well bound
heavy ions, were based on the observation that at sub-barrier
energies the measured fusion cross sections were considerably
larger than those predicted by simple barrier-penetration type
calculations using ``standard'' potentials that well described
fusion above the barrier, see e.g.\ \cite{Bec80,Sto80}. Enhancement
or suppression of fusion is therefore defined relative to the
single barrier penetration model calculation prediction using some
``reference'' potential, derived either from a fit to the above
barrier fusion cross sections or from a double-folding type
calculation. Some more recent examples directly relevant to weakly
bound nuclei may be found in e.g.\ \cite{Rus04,Kol98,Das99}. The
second school of thought compares the fusing system of interest
either with some ``reference'' system that forms the same compound
nucleus in complete fusion or the fusion of some ``reference''
nuclide with the same target, see e.g.\ \cite{Pad02}. In this
review we shall define fusion enhancement or suppression relative
to a barrier penetration calculation with a ``bare'' potential of
double-folding type using realistic matter densities. These
densities are the critical ingredient in the definition of the
potential. Reasons for this choice and the difficulties associated
with it will be discussed below.

\end{itemize}

%We adopt the above as working definitions for the purposes of the
%current review. However, there are a number of cases where the
%exact definition of a process is ambiguous.
We mentioned that there are a number of cases where the exact
definition of a process is ambiguous. The choice of nomenclature
may be more than mere semantics, as it implies the choice of a
reaction mechanism with all that this entails when one attempts to
simulate a given process theoretically. We have already discussed
the important examples of incomplete fusion, fusion following
breakup and sequential transfer. As another example, breakup may be
thought of as a particular case of transfer, i.e.\ that of
``transfer to the continuum''. For example, the transfer process:
$^A$X($^{11}$Be, $^{10}$Be)\{$^A$X + $n$\} to the $n$ + $^A$X
continuum of $^{A + 1}$X is equivalent to (and indistinguishable
from) the breakup process: $^A$X + $^{11}$Be $\rightarrow$
$^{10}$Be + $n$ + $^A$X. See \cite{Mor06} for a full discussion of
this point. However, reactions leading to $^{A + 1}$X final states
that are distinct, narrow $n$ + $^A$X resonances are more naturally
considered as conventional transfer reactions.

These ambiguities in choice of nomenclature can lead to confusion,
as the same measured quantity may be referred to by two or more
different labels that imply different production mechanisms. We
have attempted to provide practical working definitions of
quantities which can be measured in the ways discussed in sections
\ref{subsec:techniques} and \ref{subsec:identification}; precise
theoretical definitions of quantities that cannot be unambiguously
measured are of little value. One is continually faced with the
problem that when considering fusion induced by weakly bound
nuclei those quantities that are most easily measured are most
difficult to calculate theoretically and \emph{vice versa}.

\subsection{Comparison of fusion cross sections
\label{subsec:def_comparison}}

A fundamental problem in choice of nomenclature concerns the
definition of enhancement or suppression of fusion. We have already
alluded to the two main schools of thought concerning this
question, and it is by no means obvious that the different
definitions will lead to similar conclusions. In posing the
question as to whether fusion is enhanced or suppressed it is
crucial to define the reference with respect to which the
enhancement or suppression is to be judged. Both definitions have
problems.

Taking our adopted procedure first, defining fusion enhancement or
suppression relative to a single barrier penetration calculation,
the main problem concerns the choice of reference potential. This
has often been taken to be an energy-independent Woods-Saxon form,
parameterised to fit the fusion cross sections well above the
average barrier. However, it has been shown that the parameters of
such potentials differ considerably from those obtained from fits
to elastic scattering data \cite{New04}, suggesting that the
Woods-Saxon form may be inadequate. Also, coupling effects may
still be significant even at energies well above the barrier and if
these couplings are not explicitly included in the fitting
procedure the fitted ``empirical'' bare potential will reflect this
omission. Therefore, we prefer to employ a double-folded ``bare''
potential for these comparisons, although this does introduce its
own set of problems through the choice of effective nucleon-nucleon
interaction and matter densities, particularly when exotic nuclei
are concerned.

The second approach, the comparison of the fusion excitation
function of interest with that for a ``reference'' system, either
one forming the same compound nucleus in complete fusion or one
where a ``reference'' projectile fuses with the same target, has a
different set of problems. The first concerns the removal of the
``trivial'' dependence of fusion cross section on the Coulomb
barrier height and the size of the fusion partners. For the
comparison of fusion involving similar stable, well-bound nuclei
methods of ``reducing'' the fusion cross section to remove these
geometric dependencies were evolved, see e.g.\ \cite{Bec82}. The
reduction process involves dividing the measured fusion cross
sections by the quantity $\pi R_\mathrm{B}^2$, where $R_\mathrm{B}$
is the barrier radius obtained from a fit to the fusion data at
energies well above the Coulomb barrier and plotting the result as
a function of $E_\mathrm{c.m.}/V_\mathrm{B}$, where $V_\mathrm{B}$
is the Coulomb barrier height, obtained from the same fitting
procedure as $R_\mathrm{B}$. This procedure, although inherently
model-dependent, was found to give results relatively independent
of the method used to extract $V_\mathrm{B}$ and $R_\mathrm{B}$
\cite{Bec82}. However, it may no longer be valid when applied to
fusion involving halo nuclei which have much larger radii than
normal nuclei of similar masses. A new procedure has been proposed
to overcome this difficulty \cite{Gom05} which leaves a residual
``geometrical'' dependence due to the halo intact.

The main problem with this approach for weakly-bound halo nuclei,
apart from the choice of reference system --- should one compare
with the fusion of the ``core'' nucleus with the same target or
with the fusion of two well-bound stable nuclei forming the same
compound nucleus in complete fusion --- is that if enhancement or
suppression of fusion is observed to occur there is no real clue as
to the process or processes that cause it. To take a concrete
example, suppose that the fusion of $^6$He with $^{208}$Pb at
sub-barrier energies is observed to be enhanced with respect to
that for $^4$He + $^{208}$Pb. Leaving aside the question of how
best to ``reduce'' the cross sections to arrive at this conclusion,
it is by no means clear in such a comparison whether this
enhancement has been caused by coupling to the breakup of $^6$He or
by coupling to transfer reactions such as
$^{208}$Pb($^6$He,$^5$He)$^{209}$Pb or
$^{208}$Pb($^6$He,$^4$He)$^{210}$Pb. It is mainly for this reason
that we prefer to adopt a definition of enhancement or suppression
of fusion relative to a single barrier penetration calculation with
a ``bare'' potential; one can, in principle, explicitly include
couplings to the processes of interest via the coupled-channels
technique to investigate directly their effect on fusion relative
to the reference calculation.

We have dwelt at some length on the question of a reference with
respect to which enhancement or suppression of fusion may be
defined as the use of different definitions in the literature has
been a major source of the confusion in this area. In the next
section we present a review of the experimental techniques used in
the measurement of cross sections with beams of light unstable
nuclei and the corresponding experimental results.

%\newpage

\section{Experimental investigations
\label{sec:exp}}

Experimental investigation of reactions involving halo nuclei has
only become possible in the last two decades with the availability
of radioactive ion beams. The results obtained so far are affected
by the characteristics of these beams. The first experiments
involving halo nuclei were performed with beams produced with the
in-flight separation technique \cite{Gei95},
%[{\color{red} \textsf{check reference}}],
at energies of some tens to several hundreds of MeV per nucleon; in
addition to the low intensities, these beams often contained
impurities in large quantities. In some favourable cases it was
possible to slow down these beams, reaching energies around the
Coulomb barrier \cite{Sig98a}. However, most measurements in this
energy region have been carried out since radioactive beams became
available at Isotope-Separation-On-Line (ISOL) facilities
\cite{Gei95}
%[{\color{red} \textsf{check reference}}]
like the Cyclotron Research Centre (CRC) in Louvain-la-Neuve
\cite{Ryc02}, GANIL-SPIRAL \cite{Jou93} and DUBNA-DRIBs
\cite{Gul99}. Another method employs the separation of reaction
products on a light target by means of magnetic fields produced by
solenoids, and has been successfully implemented at the TwinSol
facility at the University of Notre Dame \cite{Kol89,Lee99,Bec03a}
and more recently at the RIBRAS facility at the University of
S{\~a}o Paulo \cite{Lic03}.

The ISOL method provides pure beams of relatively good optical
quality, although the intensities achievable with radioactive
species are very low when compared to stable beams. Average
intensities at the facilities mentioned above are of the order of
10$^5$-10$^7$ particles per second (pps). This fact has several
consequences for the quality of the measurements.

A first obvious problem is the limited precision of measurements,
since often only poor statistics can be collected. Another problem
concerns the measurement of the beam intensity: on the one hand the
 currents are too low to be measured in a conventional Faraday cup,
while on the other hand it is not possible to count the individual
particles in beam detectors since the latter are very soon limited
by saturation. The beam intensity is one of the factors that
determine the data normalisation, together with the target
thickness and the efficiency and accuracy of the detection method.
Systematic errors can therefore be introduced, and special care
needs to be taken to minimise this possibility.

If the elastic scattering is measured, a normalisation for the beam
charge and target thickness can be derived from the forward-angle
data, where the elastic cross section at energies around the
Coulomb barrier follows the Rutherford formula. However, if the
interest lies in other reaction channels the detection setup will
not be optimised for elastic scattering; in addition, the
efficiency of the detection will be different for elastic
scattering and the process of interest and this has to be taken
into account.

For fusion, one common practice has been to measure the cross
section up to energies sufficiently high above the barrier that
most calculations predict essentially identical fusion cross
sections and to normalise the experimental results to these
theoretical values. Another possibility is to measure, using the
same setup and in the same conditions as with the radioactive
isotope (weak beam intensity), the cross section for a nearby
stable nucleus for which the above barrier fusion cross section may
be calculated with more confidence. Conclusions have also been
derived in some cases from a direct comparison of the experimental
results for unstable and stable nuclei (see however the discussion
in section~\ref{subsec:def_comparison}).

A further aspect is the measurement of excitation functions (cross
section values at several beam energies). With weak radioactive
beams the number of different energies where measurements are
carried out is limited by the time necessary to collect sufficient
statistics. In addition, radioactive beams produced with the ISOL
method are usually post-accelerated by cyclotrons as this provides
the best possibility for beam purification (this is the case at
CRC, SPIRAL, and DRIBs); however, changing the final beam energy
takes considerably longer with a cyclotron than, for example, a
linear accelerator. To partially compensate for the weak beam
intensity thicker targets may be used; however, the corresponding
energy loss of the beam in the target should be limited (and
well-known) to retain a good definition of the actual energy of the
projectiles inducing the reaction.

For the reasons listed above, it is clear that the quality of data
collected using unstable beams cannot approach that obtained with
intense stable beams. Various methods, which we describe in section
\ref{subsec:techniques}, have nevertheless been used to tackle the
different problems.

Directly connected to the detection method is the issue of the
identification of the reaction process: using the list presented
and discussed in section~\ref{sec:definitions} as reference, we
shall see that, experimentally, it is difficult to separate the
contributions from different processes and a comparison with
calculated cross sections can only be made with great care. We come
back to this issue in section~\ref{subsec:identification}, after a
discussion of the experimental techniques.

\subsection{\it Techniques
\label{subsec:techniques}}

Due to the low intensity of radioactive ion beams, experimental
methods have necessarily focused on the efficiency of the detection
setup: mostly, large detector arrays have been employed to maximize
angular coverage, in combination with efficient signatures for the
process of interest.

The measurement of elastic scattering is facilitated by its being
the predominant process at energies around the barrier, at least at
forward angles. The integral of the elastically scattered particles
is usually easy to evaluate. However, good angular resolution is
necessary for the measurement of an angular distribution. To
measure breakup, particle identification is necessary. This
requires a more sophisticated setup, and only a few measurements
have so far been performed with exotic nuclei.

For fusion processes the choice of the target is crucial to ensure
an efficient signature. The latter is determined by the way the
compound nucleus formed in the fusion process de-excites; however,
it is rarely exclusive of a particular reaction process, so that
inclusive cross sections are often measured. This fact represents a
major problem and has eventually been the source of sometimes
contradictory conclusions. The latest experiments address this
issue, attempting to measure exclusive cross sections for a
particular system.

The way compound nuclei de-excite depends upon their mass; since we
consider here only light projectiles with masses up to $A=11$, the
detection method is determined by the choice of the target.
De-excitation may take place by charged-particle or neutron
evaporation, or by fission for the heaviest systems. The different
possibilities may also be present together. Evaporated particles
cannot be used to unequivocally identify the reaction channel, thus
their detection is not crucial. In turn, the excited residue can be
identified through its direct $\gamma$-ray emission and/or, when
its ground state is unstable, through the detection of the
characteristic decay radiation. In the following we discuss the
various possibilities in more detail.

\subsubsection{\it Identification of evaporation residues
\label{subsubsec:residues}}

After the possible evaporation of neutrons and/or charged
particles, residues may emit different kinds of radiation (mostly
depending on their mass) with different half-lives; this determines
the setup used for the identification. From the number of detected
decays (or decay chains), the cross sections in all evaporation
channels have to be determined, then summed together to obtain the
fusion cross section. If some channels are not measured, it may be
necessary to rely on evaporation codes to extrapolate the
respective cross sections; this is often the case when a residue is
stable against $\beta$- or $\alpha$-emission.

The activity per unit mass induced in the target is a combination
of the irradiation time and the half-life of the residue. Short
half-lives (up to a few seconds) require an on-line setup and a
pulsed beam, with periods of the same order of magnitude as the
half-life. Separate runs may be required to measure different
isotopes, thus increasing the total time needed for a measurement.
If the half-life is longer (in excess of some minutes at least, but
not longer than a few months at maximum), an off-line setup can be
used, which is then free from the constraints imposed by the
conditions around the target.

In both cases, normalisation can be difficult. Besides a good
calibration of the detection efficiency it requires a
\emph{continuous} measurement of the beam intensity, to correctly
determine the production rate of short-lived isotopes. Further
aspects are discussed below for the various types of radiation.

\begin{itemize}

\item \emph{Detection of $\alpha$ activity}. By using targets
around the \nuc{208}{Pb} region, the evaporation residues decay via
emission of $\alpha$ particles. Charged-particle detectors are used
to measure the energies; their characteristic values can be used,
together with the half-lives, to determine precisely the number of
decays of each isotope. Correlations in decay chains help in the
evaluation. The different half-lives of the isotopes in the various
channels may require both on-line and off-line measurements. This
method has been applied to \nuc{6}{He} (see
section~\ref{subsubsec:kolata}) and \nuc{11}{Be}
(section~\ref{subsubsec:signorini}) induced fusion.

\item \emph{Detection of $X$-ray activity}. With an appropriate
choice of target (stable, but proton-rich), the evaporation
residues are unstable against $\beta^{+}$-decay and electron
capture (E.C.). If the latter process takes place, $X$-rays are
emitted in the rearrangement of electrons in the shells; they are
characteristic of the element and thus identify the charge number
$Z$. The contributions of different isotopes can, in favourable
cases, be disentangled by following the time behaviour of an
$X$-ray line and fitting it using the known half-lives. These
half-lives are mostly of the order of minutes or longer.
Appropriate Si(Li) detectors can be used in an off-line setup,
where $X$-rays can be detected with 100 \% intrinsic efficiency and
extremely low background. The normalisation has to take into
account the self-absorption of $X$-rays in the target; in addition,
the geometrical efficiency of detection can be sensitive to the
distribution of the activated nuclei in the targets. This technique
was used for the first time with a radioactive beam in the
measurement of \nuc{6}{He} on \nuc{64}{Zn}, see
section~\ref{subsubsec:alessia}.

\item \emph{Detection of in-beam $\gamma$-rays}. The
characteristics of this method have been described in \cite{Gom89}.
The residues are identified through the emission of $\gamma$-rays
corresponding to transitions between states. The main advantage of
this technique is the fact that a residue does not need to be
unstable against $\beta$- or $\alpha$-decay; all evaporation
channels can be directly investigated. To determine the total
number of residues produced in a certain channel, all transitions
from low-lying states are usually considered; extrapolations may be
necessary to reconstruct the population of the $J=0$ ground state
when the energy of the transition is too low in even-even nuclei. A
correction for the direct production of residues in their ground
state may also have to be taken into account. However, usually the
target nuclei are chosen such that this correction is negligible.
To properly reconstruct the decay patterns a solid knowledge of the
decay scheme of all the nuclei involved is necessary. The method
requires good resolution for the detection of $\gamma$-rays in
order to correctly identify the transitions, thus the use of HPGe
detectors. With radioactive beams where high efficiency is also
important, the only measurement so far carried out employed the
EXOGAM detector array at GANIL \cite{Nav04}, where the residues
produced in \nuc{6}{He} + \nuc{63,65}{Cu}, \nuc{8}{He} +
\nuc{63}{Cu} and \nuc{6}{He} + \nuc{188,190,192}{Os} were measured
at energies slightly above the barrier. An important issue for
$\gamma$-ray measurements is background: for its reduction, a
coincidence trigger is normally used, provided either by the beam
accelerator or by ancillary detectors; the latter can be
charged-particle detectors and measure direct reaction cross
sections simultaneously.

\item \emph{Direct detection of evaporation residues}. If the
target nuclei are of light or medium-mass, the residues may have
enough forward momentum to escape the target, making direct
detection possible. The residues can then be identified in
charged-particle $\Delta E$--$E$ telescopes, in magnetic
separators, or by measuring their recoil velocity. Only the first
method has been used to date with radioactive beams: the
\nuc{13}{N} + \nuc{9}{Be} fusion cross section was measured at the
CRC \cite{Fig04}, identifying the residues in special monolithic
silicon telescope detectors \cite{Mus98}. The small velocity of the
recoils and the large momentum dispersion are the main obstacle for
the use of magnetic separators behind the target.
%[{\color{red} \textsf{Two lines about the velocity
%measurements?}}]

\end{itemize}

\subsubsection{\it Fission
\label{subsubsec:fission}}

If the compound nucleus has a large mass and is produced with an
excitation energy approaching its fission barrier \cite{Van73},
fission becomes a possible de-excitation mode. The branching ratio
is at maximum a few percent for systems like \nuc{6,7}{Li} +
\nuc{209}{Bi} at energies about 1.5 times the potential barrier. By
using a fissile target like \nuc{238}{U}, however, the fission
barrier of compound nuclei is at an energy around
$V_\mathrm{FB}\simeq6$~MeV only, and the fission probability
becomes $\sim$100 \% for projectile energies close to the Coulomb
barrier.

The two massive fission fragments are emitted back-to-back in the
centre-of-mass reference frame; for light projectiles this
approximately coincides with the laboratory frame. The large energy
of the fragments (a few tens to a hundred MeV) allows them to
escape the target foil; charged-particle detectors, covering a
large fraction of the total solid angle around the target, are used
to identify the fragments based on their energy and, possibly, by
the coincident detection of both particles. The setup needs to be
calibrated with a fission source to determine its geometrical
efficiency, and a correction may have to be introduced to account
for the forward momentum of the compound nucleus.

The fission cross section obtained in this way is an inclusive one,
as fission can be induced not only by (complete and incomplete)
fusion but also by direct processes like inelastic scattering and
transfer, leading to an excitation of the residual nucleus above
the fission barrier. It is then necessary to devise a means of
separating the possible contributions if one wishes to draw
conclusions about the fusion process.

\subsection{\it Identification of reaction processes
\label{subsec:identification}}

We introduced and discussed in section~\ref{sec:definitions}
definitions of different reaction processes. These definitions are
meant to provide a basis for comparison between different
measurements and between measured and calculated quantities. We
have already discussed some of the problems faced by calculations
due to the different possibilities for simulating a process
theoretically. Here we discuss the difficulties present on the
experimental side.

%Theoretical models can predict cross sections for particular
%processes, under assumptions which are tested by measurements. The
%comparison between calculations and experimental results should
%only be made if both quantities refer to the same reaction
%mechanism. In reality, the difficulties on both sides --- choice of
%models and definition of reaction signatures --- make this task very
%difficult. In the past the problem has sometimes been
%underestimated, and conclusions were drawn from comparisons of
%quantities referring to (partially) different processes: this was
%for example the case with \emph{complete} and \emph{total} fusion.

Experimentally, the identification of a reaction process is
provided by a \emph{signature} (detection of a particle, radiation,
coincident detection, and so on). The corresponding cross section
may be \emph{exclusive} if the signature is produced by a single
process; otherwise we speak about an \emph{inclusive} cross
section. The characteristics of weakly bound nuclei, and in
particular of halo nuclei, may be responsible for modifications to
the probability of a particular process with respect to the
behaviour of more strongly bound systems. If different processes
are measured together in an inclusive cross section the differences
may be washed out and the peculiar effects overlooked; the effect
of a particular reaction mechanism could be erroneously attributed
to a different process. It is therefore important to measure
exclusive cross sections, even if this has only been possible in a
few cases so far.

Elastic breakup (``breakup'' according to our definition) is
identified by the coincident detection of the fragments; the cross
section can be obtained in a straightforward way, taking into
account a correction for the detection efficiency. Care has to be
taken in order to unambiguously identify the fragments.

For fusion events, the signatures used for identification are those
related to the detection of evaporation residues or fission, as
discussed respectively in sections \ref{subsubsec:residues} and
\ref{subsubsec:fission}. However, either method alone is not
sufficient to separate different types of fusion; in fact these
signatures are not even exclusive of total fusion, they rather
provide inclusive cross sections as discussed below. Access to
exclusive cross sections is possible in some cases through a
careful analysis supported by theoretical calculations, in other
cases only if further detection means are available to add
information on the process. We explore the different cases below.

If the de-excitation mode of the compound nucleus is through
evaporation, in principle it is possible to determine the
\emph{complete fusion} cross section by summing all the
contributions from the different evaporation channels; that is, by
counting the corresponding residual nuclei in each channel.
However, the same nuclei may be produced by different processes,
like transfer or incomplete fusion. This is particularly true when
the target is a medium-mass nucleus, since evaporation of charged
particles is important in that case. For heavy targets (for example
Os, Bi, Pb) the main de-excitation mode is through evaporation of
neutrons; however, identification may have to be based upon
detection of daughters from the $\alpha$-decay of the original
residue, thus presenting the same problem.

A number of methods can be used to help evaluate which part of the
detected production of a residual nucleus is due to evaporation
from a particular compound system. The predictions of evaporation
codes give a reliable estimate of the ratio between the different
evaporation channels from a compound nucleus with a given
excitation energy: if a different process gives an additional
contribution to a particular channel, the comparison with such
predictions will help to identify the discrepancy and correct for
it. An additional contribution, if present, can also be recognized
if the nucleus belongs to a decay chain where the production of the
mother nucleus is also measured. Another technique has been applied
to nuclei decaying by $\alpha$-particle emission in \cite{Das04}:
the analysis of the shape of the peaks in the $\alpha$ spectrum
revealed at which depth the nucleus was implanted in the catcher
foil, thus giving information about its original energy and, in
turn, on the production process. Further methods require the
coincident detection of more observables. Particles identified as
emitted at the moment of the reaction can rule out complete fusion,
if a safe separation between reaction products and evaporated
particles can be achieved (for example, if the reaction products
present a distinctive angular distribution). Information on
$\gamma$-ray multiplicity (obtained using an efficient detector,
for example a BGO array) gives an estimate of the average angular
momentum \cite{Wuo86}, which can then be used to predict the
relative population of the different evaporation channels
\cite{Das91}.

% PRL 88 172701: the average angular momentum of a fusion reaction
%can be obtained by gamma multiplicity and by the excitation of the
%compound nucleus evaluated through the ratio of population of
%channels; it is related to the fusion excitation function

Using the techniques described above, usually in combination, it is
possible to obtain a measurement of the cross section for complete
fusion\footnote{In fact, the quantity defined in this way includes
the fusion of \emph{all} fragments following breakup, a process
that cannot be distinguished experimentally from complete fusion,
as we mentioned in section~\ref{sec:definitions}.}. The cross
section corresponding to the additional production of some residual
nuclei is then attributed to other processes, like incomplete
fusion and transfer reactions.

If the compound nucleus formed in the fusion reaction de-excites
through fission, complete fusion events can be identified as those
for which the two fission fragments are the only emitted particles.
If other processes are capable of producing fission, separation is
possible through the detection of the light quasi-projectile
fragment that emerges from the reaction.

Further identification of these processes (incomplete fusion vs.
transfer) is not trivial, and requires information about the status
of the heavy product. For example, events where the latter is
formed in a narrow resonant state may be attributed to transfer. As
mentioned in section \ref{sec:definitions} when discussing
``incomplete fusion'', it could be argued that in the case of light
projectiles at energies around the Coulomb barrier a separation may
not even be significant experimentally: the energetics of the two
mechanisms are similar, as are the angular momenta involved (which
amount to a few units only), leading to the same range of
excitation energies for the heavy product. We have already pointed
out that attempts at achieving a separation have nevertheless been
reported for the system \nuc{7}{Li} + \nuc{165}{Ho} \cite{Tri05}.

\subsection{\it Results of measurements
\label{subsec:exp_results}}

In this section we present the set of experimental results that we
use for our discussion. It includes the data collected so far on
light unstable nuclei, for which the fusion cross section around
the Coulomb barrier has been measured. Results for ``companion''
stable nuclei, obtained in the same measurements, are also
included. The data are ordered by increasing mass of the projectile
and target, rather than in a chronological order. Our aim here is
to describe what was reported by the authors in their work, along
with their conclusions; to facilitate this task, original figures
are used. We will undertake a comparative discussion of all the
systems and draw general conclusions in
section~\ref{sec:discussion}.

A more comprehensive compilation of recent results, also including
stable nuclei, has been presented in a recent review by Canto
\emph{et al.\/} \cite{Can06}.

%Data are shown in a uniform fashion, with the cross section plotted
%as function of both the centre-of-mass energy $E_\mathrm{c.m.}$ and
%the ratio $E_\mathrm{c.m.}/V_\mathrm{B}$ with respect to the
%potential barrier (as calculated in section~\ref{sec:discussion},
%see Table~2).
%In addition, we made a selection of results obtained on weakly
%bound stable light nuclei like \nuc{6}{Li}, \nuc{7}{Li} and
%\nuc{9}{Be}. The choice of the latter data is somewhat arbitrary,
%as many different measurements have been performed. Our selection
%was mainly guided by the quality of the data and their
%completeness, choosing where possible systems that provided data on
%the fusion and direct reaction cross sections. Finally we also
%include recent data on the unstable, but not-halo, nucleus
%\nuc{7}{Be}.

\subsubsection{\it \nuc{6}{He} + \nuc{64}{Zn}
\label{subsubsec:alessia}}

The \nuc{6}{He} beam available at the CRC in Louvain-la-Neuve
\cite{Ryc02} was used in a series of experiments to measure elastic
scattering, direct reaction \cite{DiP03} and fusion \cite{DiP04}
cross sections on a \nuc{64}{Zn} target. The same quantities were
also measured using a \nuc{4}{He} beam in the same energy region.

\begin{figure}[tb]
  \begin{center}
%    \begin{minipage}[t]{8 cm}
      %\epsfig{file=alessia.eps,scale=0.5}
      \includegraphics[width=.9\textwidth]{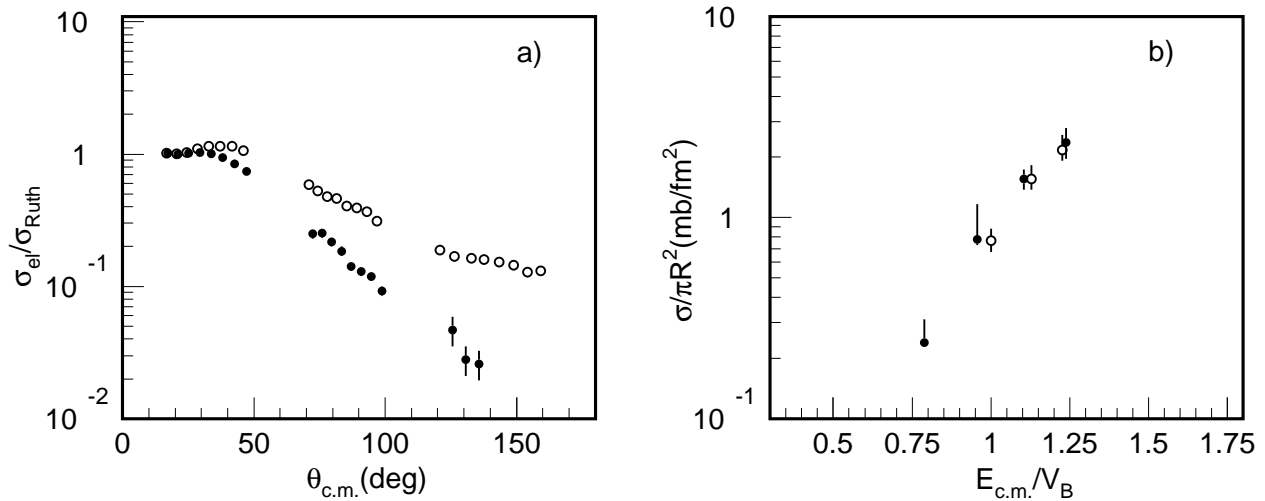}
%      \end{minipage}
%    \begin{minipage}[t]{16.5 cm}
      \caption{Results of the measurements for
      \nuc{4}{He} + \nuc{64}{Zn} (open circles) and
      \nuc{6}{He} + \nuc{64}{Zn} (filled circles): a) elastic
      scattering at $E_\mathrm{c.m.}=12.4$~MeV: the smaller elastic
      cross section for \nuc{6}{He} corresponds to a total
      reaction cross section a factor of 2.2 larger;
       b) fusion cross
      section normalised by the geometrical factor $\pi R^2$,
      plotted as a function of $E_\mathrm{c.m.}/V_\mathrm{B}$. The
      values for \nuc{6}{He} + \nuc{64}{Zn} have been corrected for
      the extra yield in the 1$\alpha$-1n evaporation channel.
      Figure adapted from \cite{DiP04}.
      \label{fig:alessia}}
%    \end{minipage}
  \end{center}
\end{figure}

Elastic scattering was measured with very good angular resolution
over a large angular range, at $E_\mathrm{c.m.}=9$~MeV for
\nuc{6}{He} and at $E_\mathrm{c.m.}=12.4$~MeV for both \nuc{6}{He}
and \nuc{4}{He}; the latter data are shown in
Fig.~\ref{fig:alessia}a. The total reaction cross sections were
extracted from the data in the three cases. At
$E_\mathrm{c.m.}=12.4$~MeV the value for \nuc{6}{He} + \nuc{64}{Zn}
is larger than that for \nuc{4}{He} + \nuc{64}{Zn} by a factor of
2.2 (see Fig.~\ref{fig:alessia}a). The largest part of the cross
section was accounted for by events where an $\alpha$ particle was
emitted: $(79\pm29)$\% at $E_\mathrm{c.m.}=9$~MeV and $(83\pm13)$\%
at $E_\mathrm{c.m.}=12.4$~MeV. An analysis of the angular
distributions of the $\alpha$ particles from \nuc{6}{He} showed
that most such events were due to direct reactions. From the
information collected it was not possible to further distinguish
between two-neutron transfer, one-neutron transfer and breakup
reactions; however, an analysis of events where two charged
particles were detected in coincidence showed that two-neutron
transfer events were certainly present.

For the measurement of the fusion cross section the activation
technique was used, with the detection of delayed $X$-ray activity
emitted by the evaporation residues (see
section~\ref{subsubsec:residues}). A stack of \nuc{64}{Zn} targets
alternated with \nuc{93}{Nb} catcher foils was used; the latter
increased the energy loss of the beam, differentiating the average
irradiation energy for each Zn target. The residues in the various
evaporation channels have half-lives between the hour and several
months. Particular care was taken in order to control the
uncertainties in the normalisation factors (detection efficiency,
beam irradiation). The fusion cross section could be obtained by
summing the yields of all evaporation channels; however, as
discussed in the previous section, the same nuclei can be produced
by fusion-evaporation and by direct reactions. In this case a
comparison of the measured ratios between the different channels
and a statistical model calculation showed a large discrepancy for
the \nuc{65}{Zn} residue (1$\alpha$-1n evaporation channel). The
excess in the yield measured for this channel was attributed to
one- and two-neutron transfer reactions; the remainder is the
\emph{complete} fusion cross section.

Di~Pietro \emph{et al.\/} chose to compare the fusion cross
sections for the \nuc{6}{He} + \nuc{64}{Zn} and \nuc{4}{He} +
\nuc{64}{Zn} systems, plotted as a function of the ratio
$E_\mathrm{c.m.}/V_\mathrm{B}$ and with the data points normalised
by the geometrical factor $\pi R^2$, where $R$ is the sum of the
projectile and target radii. Fig.~\ref{fig:alessia}b is reproduced
from the original paper. The authors observed that the excitation
functions are very similar and concluded that no sub-barrier
enhancement of the fusion cross section is seen for \nuc{6}{He}.
They underlined instead the importance of direct channels, and
remarked how the latter give a strong contribution in one
particular evaporation channel, resulting in an enhancement of the
total reaction cross section that could be misinterpreted as a
fusion enhancement".

A new measurement has recently been performed \cite{DiP06}, which
extends the energy range of the excitation functions towards higher
energies.

\subsubsection{\it \nuc{6}{He} + \nuc{206}{Pb}
\label{subsubsec:penion}}

Yu.\ Penionzhkevich \emph{et al.\/} \cite{Pen06} measured the
\nuc{6}{He} + \nuc{206}{Pb} system at the recently commissioned
ISOL facility DRIBs \cite{Gul99} at the Flerov Laboratory in Dubna.
The initial energy of the \nuc{6}{He} beam was
$E_\mathrm{lab}=60.3$~MeV, thus degraders had to be employed to
reach the energy range around the Coulomb barrier
($V_\mathrm{B}\simeq19$~MeV), resulting in a very large energy
spread of the incident particles. A stack of \nuc{206}{Pb} targets
was used, which allowed simultaneous measurements at different
energies between 23~MeV and 13~MeV.

\begin{figure}[tb]
  \begin{center}
%    \begin{minipage}[t]{8 cm}
      %\epsfig{file=penion.eps,scale=0.5}
      \includegraphics[width=.6\textwidth]{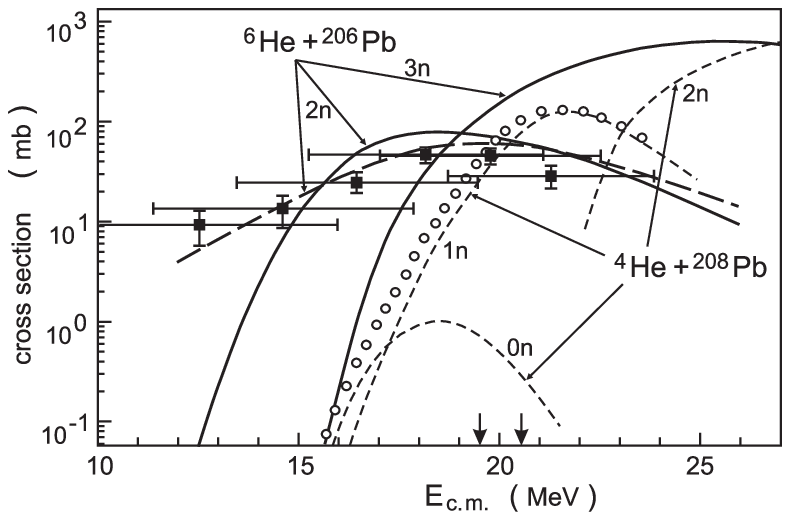}
%    \end{minipage}
%    \begin{minipage}[t]{16.5 cm}
      \caption{(taken from \cite{Pen06}) Cross sections for
      the production of \nuc{210}{Po} in the
      \nuc{6}{He} + \nuc{206}{Pb} reaction (solid squares) and for
      the production of \nuc{211}{Po} in the
      \nuc{4}{He} + \nuc{208}{Pb} reaction (circles). Dashed lines
      are predictions from a statistical model for the evaporation
      channels populated in \nuc{4}{He} + \nuc{208}{Pb}. Solid
      lines are predictions from~\cite{Zag03}; long-dashed lines
      take into account the energy spread. The arrows on the energy
      axis indicate the position of the Coulomb barrier for
      \nuc{6}{He} (left) and \nuc{4}{He} (right).
      \label{fig:penion}}
%     \end{minipage}
  \end{center}
\end{figure}

Results were reported for the two-neutron evaporation channel,
measured using the activation method. The residual nucleus
\nuc{210}{Po} decays emitting characteristic $\alpha$ particles at
$E_\alpha= 5.3$~MeV with a half-life $T_{1/2}=138$~d. The measured
cross section data were compared with those for the 1n-evaporation
channel in \nuc{4}{He} + \nuc{208}{Pb}; both are shown in
Fig.~\ref{fig:penion}. The authors pointed to the large difference
in the behaviour of the two cross sections at energies below the
Coulomb barrier and concluded that there is a \emph{large
enhancement} of the fusion cross section for \nuc{6}{He} +
\nuc{206}{Pb}. In a model of ``sequential fusion''
%\footnote{However, this is essentially a two-step process, for
%which we refer to the considerations made in the Introduction.}
proposed by Zagrebaev~\cite{Zag03,Zag04}, the transfer of the two
``valence'' neutrons in $^6$He (with positive $Q$-value) would
cause a gain in the energy of relative motion of the two nuclei and
thus facilitate fusion: such a prediction was found to agree with
the reported data (Fig.~\ref{fig:penion}).

These experimental results are particularly surprising, as they
show that at deep sub-barrier energies a sizable cross section is
found. However, the large energy spread in these data, and the
absence of a reference measured with the same experimental setup
suggest that further investigation is necessary.

\subsubsection{\it \nuc{6}{He} + \nuc{209}{Bi}
\label{subsubsec:kolata}}

The reaction channels in the \nuc{6}{He} + \nuc{209}{Bi} system
were studied in a series of measurements at the TwinSol facility at
the University of Notre Dame~\cite{Kol89,Lee99,Bec03a}. Two
superconducting solenoids were used to separate and focus the
nuclei produced in reactions on thin targets. For \nuc{6}{He}, a
\nuc{7}{Li} primary beam was used with a \nuc{9}{Be} target. One
concern of this technique is represented by contaminants in the
beam, which have to be eliminated or identified; various methods
were developed in the course of the measurement campaign to reach
this goal.

The set of cross sections reported by the Notre Dame group is the
most complete for this kind of system. It includes
fusion-fission~\cite{Kol98a}, the 4n evaporation
channel~\cite{DeY98}, the 3n channel~\cite{Kol98}, elastic
scattering~\cite{Agu01}, the total \nuc{4}{He}-production at
near-barrier~\cite{Agu00} and sub-barrier~\cite{Agu01} energies,
the 1n-~\cite{Byc04} and 2n-transfer channels~\cite{DeY05}.

\begin{figure}[tb]
  \centering
    \begin{minipage}[t]{0.98\textwidth}
      \begin{center}
      \includegraphics[width=.48\textwidth]{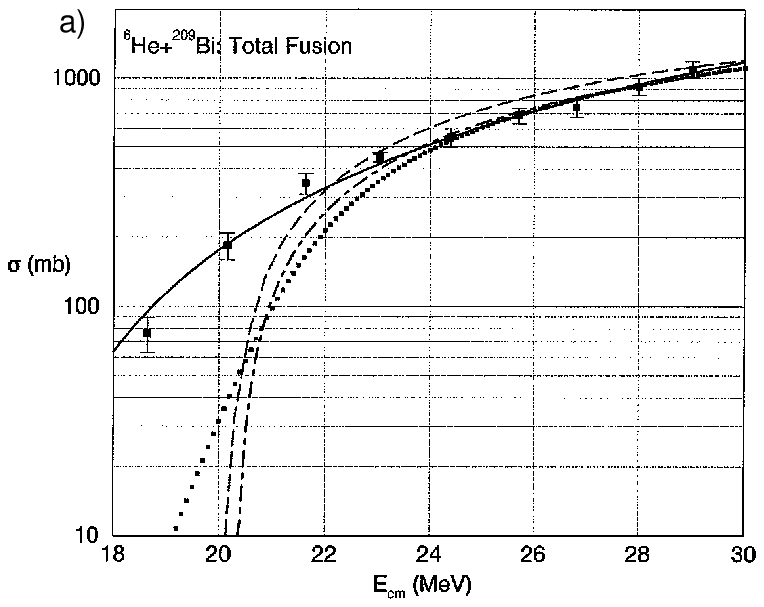}
      \includegraphics[width=.48\textwidth]{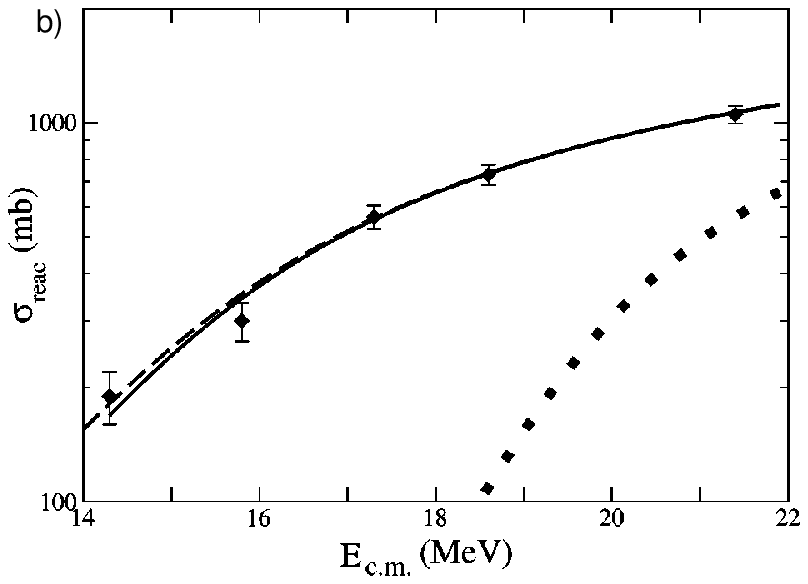}
      \end{center}
      \end{minipage}
%    \begin{minipage}[t]{16.5 cm}
      \caption{a) (from~\cite{Kol98}) \nuc{6}{He} + \nuc{209}{Bi}
      total fusion cross section; the dashed curve is the
      prediction of a statistical model, the dotted curve is a pure
      barrier penetration calculation, the dot-dashed curve is a
      $1/E_\mathrm{c.m.}$ fit to the high-energy data and the solid
      curve is a fit according to the Stelson model \cite{Ste90}; b)
      (from~\cite{Agu01}) total reaction cross section for
      \nuc{6}{He} + \nuc{209}{Bi}; the solid and dashed curves are
      calculations using parameters obtained from optical model
      fits to the elastic scattering data; the dotted curve is a
      prediction for the well-bound nucleus \nuc{7}{Li} on
      \nuc{208}{Pb}.
      \label{fig:kolata}}
%    \end{minipage}
\end{figure}

For fusion at energies in the vicinity of the potential barrier,
the 3n and 4n evaporation channels are the most important. They
were measured using the activation technique, detecting the
$\alpha$ particles emitted in the on-line and off-line decay (3n
and 4n channel, respectively) of the evaporation residues.
In~\cite{Kol98} the sum of the two cross sections is compared to a
statistical model calculation and to a simple barrier penetration
calculation, resulting in the plot reported in
Fig.~\ref{fig:kolata}a. The plot also presents the result of a fit
meant to determine the position of an effective barrier according
to the ``neutron flow'' phenomenological model of Stelson \emph{et
al.\/} \cite{Ste90}. From their results, the authors concluded that
there is an enhancement of the \emph{total} fusion cross section at
energies below the barrier and noted the good fit provided by the
Stelson model; the model is related to the low-binding energy of
neutrons in the projectile and thus transfer channels with positive
$Q$-values, as in the case of the \nuc{6}{He} + \nuc{209}{Bi}
system.

Of particular interest is the observation of a large yield for
$\alpha$ particles~\cite{Agu00}, exhausting the total reaction
cross section resulting from optical model fits to the elastic
scattering~\cite{Agu01} (similar to that observed for \nuc{4}{He}
+ \nuc{64}{Zn}). The total reaction cross section is shown in
Fig.~\ref{fig:kolata}b. In~\cite{DeY05} at least 55\% of this yield
is attributed to the 2n-transfer channel to unbound states in
\nuc{211}{Bi}.

It has to be mentioned that the \nuc{6}{He} + \nuc{209}{Bi} system
has also been measured at the Flerov laboratory in Dubna at
energies above the potential barrier ($E_\mathrm{c.m.}=25$ to
30~MeV), using a \nuc{6}{He} beam produced with the in-flight
separation technique. Cross sections for fission and the 4n
evaporation channel were reported~\cite{Pen95a,Fom95,Pen02}; the
former was found to be a factor 3 to 4 larger than the
corresponding \nuc{4}{He} + \nuc{209}{Bi} cross section, however
the large uncertainties at the lower energies ($E_\mathrm{c.m.}$
smaller than 35~MeV) prevented firm conclusions. These results were
not confirmed by the measurement of Kolata et al.~\cite{Kol98a},
where the cross section at the same energy was reported to be about
an order of magnitude smaller.

\subsubsection{\it \nuc{6}{He} + \nuc{238}{U}
\label{subsubsec:6he238u}}

In these measurements, fission was chosen as the signature to
identify fusion. The nuclei in the region around \nuc{238}{U} have
low fission barriers, $V_\mathrm{FB}\simeq6$~MeV \cite{Van73}. For
projectile energies around the potential barrier
($V_\mathrm{B}\simeq21$~MeV for \nuc{6}{He} and
$V_\mathrm{B}\simeq23$~MeV for \nuc{4}{He}), a fusion reaction will
always produce a compound nucleus at an excitation energy well
above the fission barrier inducing a fission probability close to
100\%.
%$E^*>V_\mathrm{FB}$, leading to de-excitation through
%fission with 100\% probability.

This measurement, like the one on \nuc{64}{Zn}, was performed using
the \nuc{6}{He} beam at the CRC in Louvain-la-Neuve. To attain a
large efficiency for the detection of the two fission fragments,
forty large-area charged-particle silicon detectors were placed on
the inner surfaces of two cubes, symmetrically arranged around the
target position, covering about 70\% of the whole solid angle.

\begin{figure}[tb]
  \begin{center}
%     \begin{minipage}[t]{0.98\textwidth}
      %\epsfig{file=raabe_6he.eps,scale=0.5}
      \includegraphics[width=.5\textwidth]{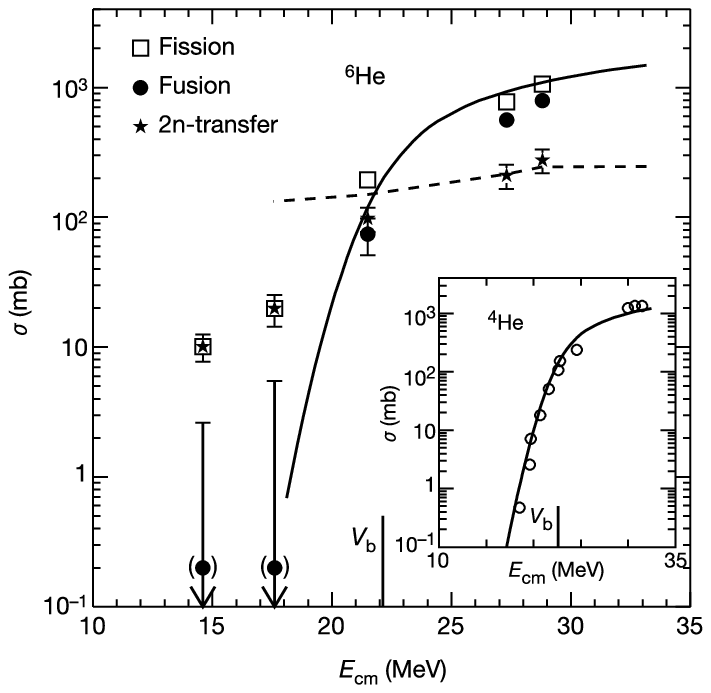}
%     \end{minipage}
%     \begin{minipage}[t]{16.5 cm}
      \caption{(from~\cite{Raa04}) Cross sections for
      \nuc{4,6}{He} + \nuc{238}{U}. The data are compared with
      calculations using an effective optical potential, which for
      \nuc{6}{He} was derived from a continuum discretised coupled
      channels (CDCC) calculation. The solid
      curves are fusion calculations using a short-ranged imaginary
      potential; the dashed line is the 2n-transfer cross section
      to excited states in \nuc{240}{U}.
      \label{fig:raabe_6he}}
%     \end{minipage}
  \end{center}
\end{figure}

As explained in section~\ref{subsubsec:fission}, other reactions
besides fusion are capable of exciting the target nucleus above its
fission barrier. In these cases, however, a quasi-projectile
fragment would be emitted in the process and, if charged, could be
detected in coincidence with the fission fragments (a correction
for the geometrical efficiency for the detection of the
quasi-projectile particle needs to be included). With beams of He
isotopes, the capture of the whole charge of the projectile would
very likely correspond to \emph{complete} fusion --- the only other
possibility, very unfavourable energetically, being for \nuc{6}{He}
the transfer of the $\alpha$ particle core. The results were
published in~\cite{Tro00,Raa04}, and they are summarised in
Fig.~\ref{fig:raabe_6he} (taken from~\cite{Raa04}). The authors
compared the behaviour of the two isotopes: the fission cross
section for \nuc{6}{He} + \nuc{238}{U} shows an enhancement at
energies below the barrier with respect to \nuc{4}{He} +
\nuc{238}{U} and also with respect to \nuc{6}{Li} + \nuc{238}{U}
(the data for \nuc{6}{Li}, from~\cite{Fre75}, are not shown in the
figure). However, for \nuc{6}{He} most of these fission events were
not due to fusion, as revealed by the detection of a
quasi-projectile particle in coincidence. Once this contribution
was subtracted, only upper limits for the complete fusion cross
section could be given for the two points measured below the
barrier, and the conclusion was that no enhancement was observed.
The process generating the large fission yield was identified
through an analysis of the energies and angular distributions of
the quasi-projectile particles to be the transfer of two neutrons
to the target to form \nuc{240}{U}. The optimum matching conditions
favour transfer to excited states of \nuc{240}{U} above the fission
barrier. Taking into account that direct breakup could not be
measured due to the required fission signature, the observation of
a large reaction cross section for direct reactions at energies
below the barrier is similar to that reported for \nuc{6}{He} +
\nuc{64}{Zn} and \nuc{6}{He} + \nuc{209}{Bi}.

\subsubsection{\it \nuc{7}{Be} + \nuc{238}{U}
\label{subsubsec:7be238u}}

\begin{figure}[tb]
  \begin{center}
%     \begin{minipage}[t]{0.98\textwidth}
      %\epsfig{file=raabe_7be.eps,scale=0.55}
      \includegraphics[width=.55\textwidth]{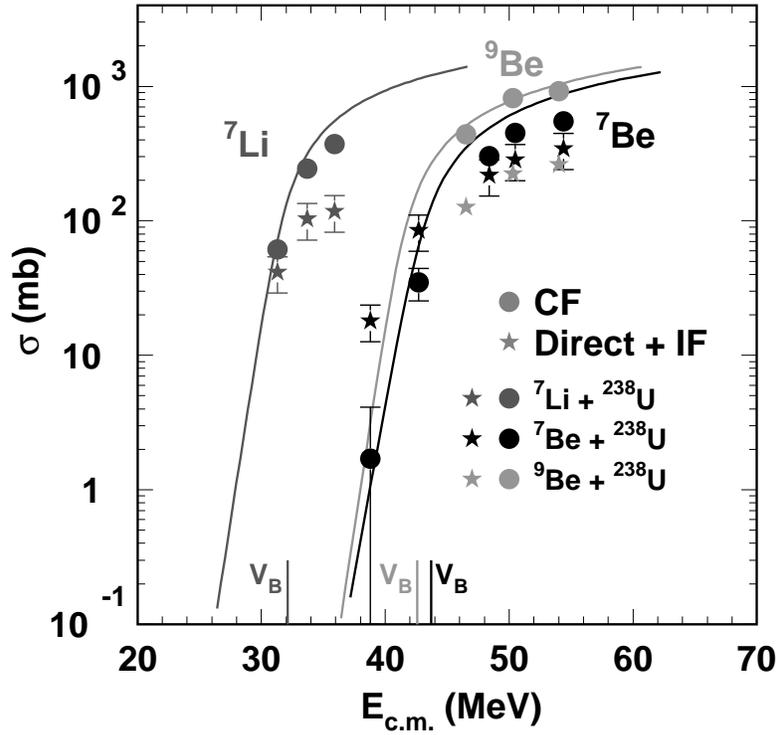}
%     \end{minipage}
%     \begin{minipage}[t]{16.5 cm}
      \caption{(from~\cite{Raa06}) Cross sections for complete
      fusion (filled circles) and incomplete fusion (IF) + direct
      processes (stars) for \nuc{7,9}{Be} and \nuc{7}{Li} on
      \nuc{238}{U}. Curves are one-dimensional barrier penetration
      model calculations for fusion.
      \label{fig:raabe_7be}}
%     \end{minipage}
  \end{center}
\end{figure}

The \nuc{7}{Be} nucleus is unstable and relatively weakly bound
(the lowest breakup threshold is that into $\alpha$ + \nuc{3}{He}
at $E^*=1.59$~MeV). However, it does not exhibit a halo structure;
we include it in this review to find possible differences with
respect to halo nuclei.

The \nuc{7}{Be} beam was produced at the CRC in Louvain-la-Neuve
from nuclei obtained from a (p,n) reaction on \nuc{7}{Li} and
chemically separated off-line (the half-life of \nuc{7}{Be} is
$T_{1/2}=53$~d). The same fission signature technique described
above was used; in the case of \nuc{7}{Be}, in addition all
possible fragments ($\alpha$, \nuc{3}{He}, \nuc{6}{Li}, d, p) are
charged and were thus detected. The \emph{complete} fusion events
can therefore be unambiguously identified as those fissions not
accompanied by the emission of light particles. The results of this
measurement appeared in~\cite{Raa06}, from which
Fig.~\ref{fig:raabe_7be} is taken. The authors compared the cross
section data for \nuc{7}{Be} with those for \nuc{7}{Li} and
\nuc{9}{Be} on the same target. Their main conclusion was that the
pattern is similar to that observed in the case of \nuc{6}{He} +
\nuc{238}{U}: at energies below the potential barrier a large total
reaction cross section is observed. However, this is mainly due to
processes other than complete fusion, since a charged light
particle was detected in coincidence. The sub-barrier complete
fusion cross section for \nuc{7}{Be} does not differ much from a
one-dimensional barrier penetration model calculation, and no
enhancement was observed. At energies around and above the barrier,
on the other hand, the fusion cross section is smaller than the
calculation, and the authors speak of a suppression of the complete
fusion. Concerning the nature of the processes generating the large
fission cross section, it was suggested that they mostly consist of
transfer of clusters to the target nucleus and possibly with a
component due to reactions characterised by a small impact
parameter and thus more properly labelled incomplete fusion.

\subsubsection{\it \nuc{10,11}{Be} + \nuc{209}{Bi}
\label{subsubsec:signorini}}

The fusion cross sections for the \nuc{10,11}{Be} + \nuc{209}{Bi}
systems were measured at RIKEN, Japan, using the target activation
method in two successive experiments ~\cite{Yos96,Sig04}. The beams
were produced using the in-flight separation technique; from an
initial energy of about 45~MeV/nucleon, the Be nuclei were slowed
down to energies around the potential barrier using thick degrader
plates. While the energy spread of the beam was very large, for
each nucleus impinging on a stack of targets the energy was
determined using the time of flight, achieving an accuracy of about
$\pm1$~MeV. Following activation, the short-lived
($T_{1/2}<1$~$\mu$s) residual nuclei in the targets decayed
emitting $\alpha$ particles. Given the short half-lives and the low
beam rate, each decay could be correlated to a projectile of known
energy, obtaining excitation functions for evaporation channels
corresponding to the residual nuclei \nuc{215,216}{Fr} (5n and 4n
evaporation channels for \nuc{11}{Be} + \nuc{209}{Bi}).

The cross section data for the weakly-bound
($E_\mathrm{B}=0.50$~MeV) one-neutron halo nucleus \nuc{11}{Be}
were compared with those for the well-bound
($E_\mathrm{B}=6.8$~MeV) \nuc{10}{Be} nucleus. In addition, the
fusion of the stable but weakly-bound ($E_\mathrm{B}=1.57$~MeV)
\nuc{9}{Be} nucleus with \nuc{209}{Bi} was measured by the same
group at the Munich Tandem accelerator~\cite{Sig99}.

\begin{figure}[tb]
  \begin{center}
%     \begin{minipage}[t]{0.98\textwidth}
      %\epsfig{file=signorini.eps,scale=0.5}
      \includegraphics[width=.5\textwidth]{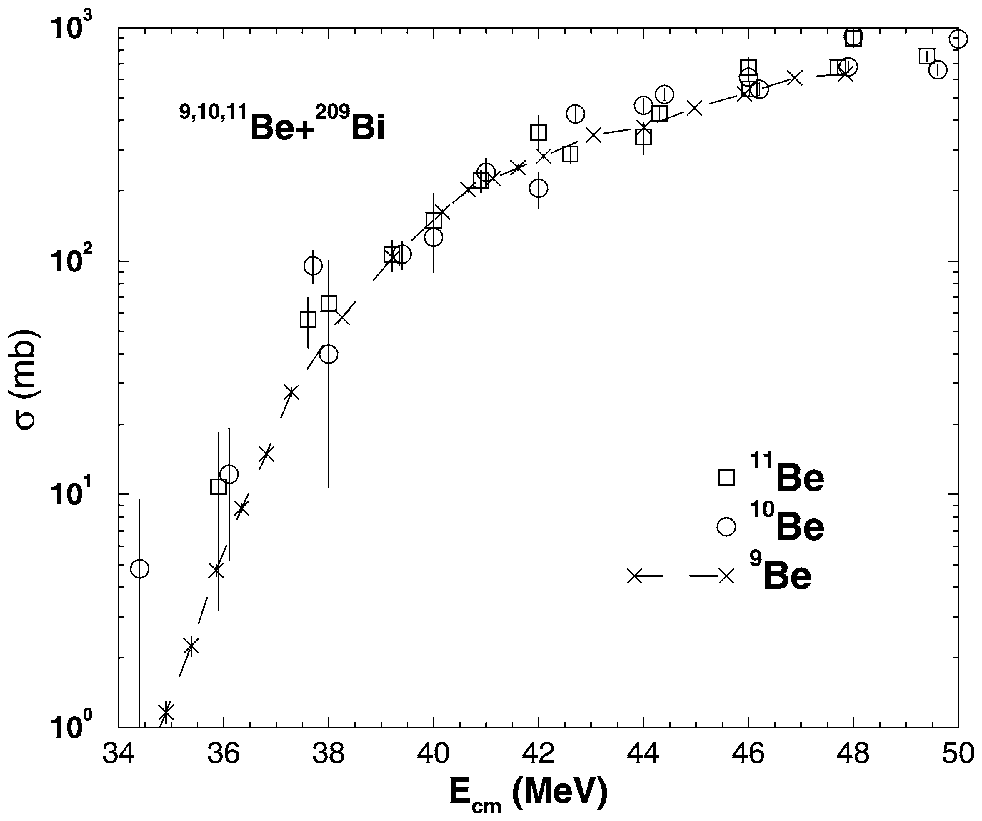}
%     \end{minipage}
%     \begin{minipage}[t]{16.5 cm}
      \caption{(from~\cite{Sig04}) Fusion cross sections for the
      systems \nuc{9,10,11}{Be} + \nuc{209}{Bi}.
      \label{fig:signorini}}
%     \end{minipage}
  \end{center}
\end{figure}

All these data are presented together in Fig.~\ref{fig:signorini},
taken from~\cite{Sig04}. Those for \nuc{11}{Be} and \nuc{10}{Be}
are total fusion cross sections, obtained as the sum of the
measured evaporation channels; the authors argue that the
incomplete fusion amounts to less than 30\% for \nuc{11}{Be} and is
negligible for \nuc{10}{Be}. For \nuc{9}{Be}, the data represent
the complete fusion cross section. In the case of \nuc{11}{Be}
(\nuc{10}{Be}), the 3n (2n) evaporation channel is not included,
which may lead to an underestimation of the cross section in the
sub-barrier region; however, the effect should be comparable for
the two systems. The main conclusion of the authors concerned the
unexpected similarity between the three cross sections, while
\nuc{11}{Be} was expected to show pronounced effects due to its
weakly-bound nature; thus, they speak of neither enhancement nor
hindrance of the fusion cross section for \nuc{11}{Be}. It was also
correctly underlined that \nuc{9}{Be}, stable but weakly bound, may
not be a good reference system for comparison~\cite{Sig01}.
\nuc{11}{Be} and \nuc{9}{Be} also show comparable behaviour
concerning elastic scattering, as recently reported by the same
group~\cite{Maz06}.

\bigskip

With the review presented above, we wish to show how experimental
investigations in this field face many challenges that may lead to
uncertain results. In addition, by reporting the original remarks
of the authors, it becomes perhaps more apparent how the sometimes
contradictory conclusions may have originated from the different
ways chosen to present those results.

In the next section we will perform  calculations and analysis for
those reaction mechanisms involving weakly bound nuclei that can be
predicted with a good degree of confidence. From the results we
will derive some conclusions about the expected behaviour of these
nuclei at energies around the potential barrier. In section
\ref{sec:discussion} we will then compare the experimental data
presented here with a simple, consistent model, and try to draw
some general conclusions about the possible effects of weak binding
and transfers.

%\newpage

\section{Theoretical models
\label{sec:theory}}

In this section we present and discuss some of the theoretical
models used to describe direct reactions and fusion at incident
energies close to the top of the Coulomb barrier. These models are
then used to investigate the effect of breakup and transfer
couplings on other reaction channels for weakly bound nuclei.

%We begin with a small ``dictionary'' of definitions for the various
%reaction processes to be discussed here, as the situation in this
%regard in the current literature is sometimes confusing.

\subsection{\it Methods
\label{subsec:methods}}

It has been firmly established that strong coupling effects are
important for both elastic scattering (\cite{Sat90} and references
therein, see \cite{Rus03,Kee03,Rus05,Mat04,Mor03,Tak03,Mat06} for
some recent examples with exotic nuclei) and fusion
\cite{Bec85,Ste86} induced by heavy ions at incident energies close
to the top of the Coulomb barrier. In particular, it was shown that
under certain approximations the single fusion barrier could be
thought of as splitting into a distribution of barriers when
couplings to other channels are present. Calculations that
explicitly include these couplings are able to satisfactorily
describe the ensemble of nuclear reaction data in this energy
regime, e.g.\ \cite{Tho85,Pie85,Kee96a,Kee98,Per06}. The standard
coupled-channels (CC) techniques for the analysis of direct nuclear
reactions are presented in a number of text books, e.g.\ Satchler
\cite{Sat83}, and will not be covered here. Similarly, there are a
number of review articles discussing the analysis of near-barrier
fusion data for stable, well-bound nuclei
\cite{Bec85,Ste86,Vaz81,Bec88,Das98}. Attempts have been made to
extend these standard techniques to enable the inclusion of breakup
for both direct reactions and fusion, essential for detailed
analyses of weakly-bound systems such as halo nuclei.

For the description of breakup and its effect on other direct
reaction channels, most importantly the elastic scattering, the
most sophisticated and successful theory currently available is
incontestably the continuum discretised coupled channels (CDCC)
method. An extension of the standard coupled channels theory to
allow treatment of couplings to the unbound continuum of states
above the breakup threshold, it was first developed to describe
deuteron breakup \cite{Raw74} (see \cite{Cha06} for a recent
systematic application to many target nuclei). The method was later
extended to other weakly-bound systems; $^6$Li \cite{Sak82}, $^7$Li
\cite{Sak86} and $^7$Be \cite{Yam89}. In particular, the
continuum--continuum coupling, i.e.\ coupling between the ``excited
states'' of the continuum, has been found to play an important
r\^ole \cite{Sak83,Sak87}. Two comprehensive reviews of CDCC
applied to deuterons \cite{Aus87} and other nuclei \cite{Sak86}
provide numerous examples of its success. The formalism has been
recently extended to include systems that break up into three
particles \cite{Mat04,Mat06,Mat04a} and to systems that include an
``active'' core \cite{Sum06,Sum06a}. These extensions are essential
for a realistic description of the breakup of weakly-bound halo
nuclei.

Although CDCC has proved remarkably successful in describing
breakup and its effect on other direct reaction channels, to date
there is no completely satisfactory way of linking it to a
calculation of fusion. A method has been proposed for calculating
\emph{total} fusion within the framework of standard CDCC
\cite{Dia03} by the use of short-ranged imaginary potentials for
each of the two fragments, simulating the ingoing-wave boundary
condition. These potentials are then folded together with the
projectile internal wave functions to produce the final imaginary
potentials. The total fusion is then defined as the amount of flux
that leaves the coupled channels set, i.e.\ the total absorption
cross section, obtained by subtracting the cross sections of all
channels explicitly included in the coupling scheme from the total
reaction cross section.
%Thus
%each time the coupling scheme is changed, e.g.\ by coupling to
%additional channels, the strict definition of the fusion cross
%section will change somewhat.
However, unfortunately this technique cannot be used to calculate
incomplete fusion as it is impossible to trace the trajectories of
the individual fragments after breakup has occurred. Also, to
describe the direct channels simultaneously with the fusion one
would need to include many other channels, e.g.\ inelastic
excitation of the target and transfers, and the couplings between
them. Such calculations are not currently tractable for most
systems, although this problem is a general one and not confined to
this particular method.

Another, much more approximate way of calculating a fusion cross
section within the CDCC framework is to employ a one-dimensional
barrier penetration model (BPM) in conjunction with a dynamic
polarisation potential (DPP) generated by the couplings to the
continuum. The BPM assumes that all the flux that penetrates the
Coulomb barrier, defined in this instance by the real part of the
effective scattering potential, leads to fusion. As the real part
of the DPP is energy dependent (particularly for energies close to
the Coulomb barrier) this model replaces the distribution of
barriers of the full coupled channels scheme with an energy
dependent one-dimensional barrier. Although this method has been
applied to fusion induced by weakly bound projectiles \cite{Kee02},
it is not completely clear whether the calculated fusion cross
sections should be compared with complete or total fusion
measurements.

There are other problems with the CDCC+DPP+BPM approach, notably in
the definition of the DPP which may be the ``trivially equivalent
local potential'' (TELP), as used in \cite{Kee02} and based on wave
functions, or a potential obtained by inversion of the S-matrix
\cite{Kuk04}. While both methods give the same elastic scattering
for a given system it is by no means clear that they will give the
same fusion cross section, as their radial forms are often very
different in the nuclear interior. It has also been found that for
well-bound nuclei the CC+DPP+BPM approach consistently
underpredicts the fusion cross sections of the more realistic
ingoing-wave boundary condition model \cite{Tho89}.

Clearly, some means of linking the CDCC description of breakup to a
realistic method of calculating fusion is required. In particular,
a method that enables the trajectories of the fragments to be
traced is needed if incomplete fusion --- in this context fusion
following breakup --- is to be reliably calculated and separated
(at least theoretically) from other processes yielding the same
residual nuclei.

%The various other techniques that have been applied to the problem
%of calculating the fusion of weakly-bound nuclei have been well
%summarised in the recent review article \cite{Can06}. However, the
%treatment of breakup in many of these methods is necessarily more
%or less approximate. What is required is some means of linking the
%CDCC description of breakup to a realistic method of calculating
%fusion, in particular one that enables the trajectories of the
%fragments to be traced so that incomplete fusion may be reliably
%calculated.

While breakup has excited the most interest among the reactions
that may be induced by halo nuclei, due to their very low
thresholds against this process, one should not overlook the
possible importance of transfer reactions. The low one or two
neutron removal energies of the halo nuclei should also favour
neutron stripping reactions \cite{Byc04,DeY05}. A reliable model
exists for the calculation of transfer processes in the coupled
reaction channels (CRC) framework, an extension of the coupled
channels formalism to include channels belonging to different
partitions. The calculation of fusion is readily incorporated
within the CRC formalism by using the ingoing-wave boundary
condition \cite{Pie85,Das83} or its equivalent \cite{Rho84}. In
this case, the standard formalism developed for well-bound stable
nuclei (see e.g.\ Satchler \cite{Sat83}) may be used without
modification for halo nuclei, although care may be needed in the
choice of bound-state form-factors in order to take account of the
halo properties.

Although we shall not discuss them further here, we note that
time-dependent wave-packet methods provide an alternative view of
the fusion problem to that provided by the coupled channels scheme.
A three-body model using this method has recently been applied to
the fusion of neutron halo nuclei with heavy targets \cite{Ito06}.

As a recent review article \cite{Can06} has already provided an
excellent summary of the considerable literature concerned with the
effect of coupling to breakup on near and sub-barrier fusion, and
owing to the lack of a theoretical model that treats fusion and
breakup equally realistically, we shall not further investigate
this subject here. However, we shall probe the influence of breakup
threshold on the elastic breakup cross section and the effect of
breakup coupling on the elastic scattering through model
calculations. We shall also investigate the effect of transfer
reactions on fusion induced by weakly bound nuclei through CRC
calculations comparing systems involving $^6$Li, $^6$He and $^8$He
projectiles. Before doing so we shall briefly list in the next
section some of the more widely available codes that are frequently
used to perform coupled channels and coupled reaction channels
calculations. Section \ref{subsec:breakup} contains an
investigation of the effect of breakup threshold energy on the
total cross section for breakup and its influence on the elastic
scattering while section \ref{subsec:transfer} concentrates on the
effect of transfer reactions.

\subsection{\it Some widely available coupled channels codes
\label{subsec:codes}}

Some of the most frequently used coupled channels codes in the
literature are \textsc{ecis} \cite{Ray81}, \textsc{ptolemy}
\cite{Rho80}, \textsc{fresco} \cite{Tho88} and \textsc{ccfull}
\cite{Hag99}. This list is not intended to be exhaustive, but
merely indicative of the types of code available for coupled
channels and coupled reaction channels calculations. \textsc{ecis},
\textsc{ptolemy} and \textsc{fresco} are general reaction codes
whereas \textsc{ccfull} calculates the fusion cross section only.
Brief descriptions of the capabilities of these codes follow.

%\begin{sffamily}
%This should be extended (Riccardo)
%\end{sffamily}

\begin{itemize}
\item \textsc{ecis} uses the full CC method to
calculate inelastic excitations of target and projectile. Transfer
reactions may be included via zero-range CRC and parameter searches
to fit a set of experimental data are possible.

\item \textsc{ptolemy}, like the previous code, calculates
inelastic excitations using full collective model CC. Transfer
reactions are included via the exact finite-range distorted-wave
Born approximation (DWBA).

\item \textsc{fresco} is a very complete code for the
calculation of direct reactions. It uses full CC for inelastic
excitations and transfer reactions are included via full
finite-range CRC (implemented iteratively) with complex remnant
term and non-orthogonality correction. Three-body breakup (i.e.\
breakup of one of the interacting partners into two fragments) may
be included via the CDCC method employing Watanabe-type cluster
folding. Parameter searches are possible using the auxiliary code
\textsc{sfresco}.

\item \textsc{ccfull} focuses on the calculation of
the fusion cross section. It can include inelastic excitations of
both particles (rotational and vibrational couplings) and a
transfer between ground states using a coupling form factor. The
effect of couplings is taken into account to all orders. The
barrier penetrability is calculated for each partial wave, and to
obtain the fusion cross section the ingoing-wave boundary condition
is used.

\end{itemize}

\subsection{\it Model calculations for breakup
\label{subsec:breakup}}

\subsubsection{\it Conditions for the breakup calculations
\label{subsubsec:breakup_ansatz}}

In this section we use the CDCC method to probe the influence of
breakup threshold and target charge, i.e.\ the magnitude of the
Coulomb potential, on the total elastic breakup cross section and
the effect of breakup coupling on the elastic scattering through a
series of model calculations using the coupled reaction channels
code \textsc{fresco} \cite{Tho88}. We employ a simplified
``$^7$Li'' projectile with ground-state spin $0 \hbar$, no resonant
states and breakup thresholds set at 1.0, 1.5, 2.0 and 2.5 MeV. As
targets, we take $^{208}$Pb, $^{58}$Ni and $^{12}$C, as
representative of systems where the Coulomb force is dominant
($^{208}$Pb) through to those where the nuclear force predominates
($^{12}$C). We employ a two-body $\alpha+t$ model of $^7$Li with
the triton spin set to $0 \hbar$ and the $\alpha+t$ continuum
discretised in the usual way \cite{Sak86} into a series of bins in
momentum ($k$) space of width $\Delta k = 0.1$ fm$^{-1}$ up to a
maximum of $k_{\mathrm{max}} = 1.0$ fm$^{-1}$. The value of
$k_{\mathrm{max}}$ was reduced for the lower incident energy
calculations as required.

All diagonal and coupling potentials were of the Watanabe
cluster-folding type as in e.g.\ \cite{Kee05}. The $\alpha$ and
triton plus target optical potentials required for the folding
procedure were taken from \cite{Hud74} and \cite{War79},
\cite{Bre94} and \cite{Sak89}, \cite{Ala90} (potential III) and
\cite{Pul64} for the $^{208}$Pb, $^{58}$Ni and $^{12}$C targets,
respectively. All these potentials are energy-independent,
therefore the energy dependencies of the calculated cross sections
are strictly dynamic, depending on the couplings alone.

\begin{figure}
\epsfysize=9.0cm
\begin{center}
  \includegraphics[width=.6\textwidth]{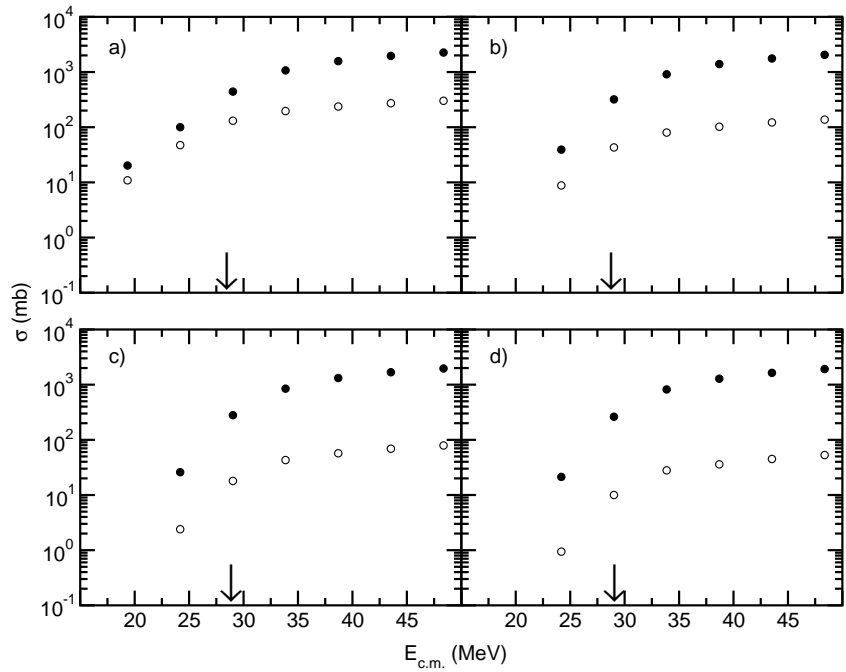}
\end{center}
\caption{Model breakup calculations for a $^{208}$Pb target with
breakup thresholds of: a) 1.0 MeV, b) 1.5 MeV, c) 2.0 MeV and d)
2.5 MeV. The filled and open circles denote the total reaction
cross section and total breakup cross section, respectively. The
arrows indicate the positions of the respective Coulomb barriers,
defined as the maximum of the combined bare nuclear plus Coulomb
potentials. \label{fig:bu208pb}}
\end{figure}
\begin{figure}
\epsfysize=9.0cm
\begin{center}
  \includegraphics[width=.6\textwidth]{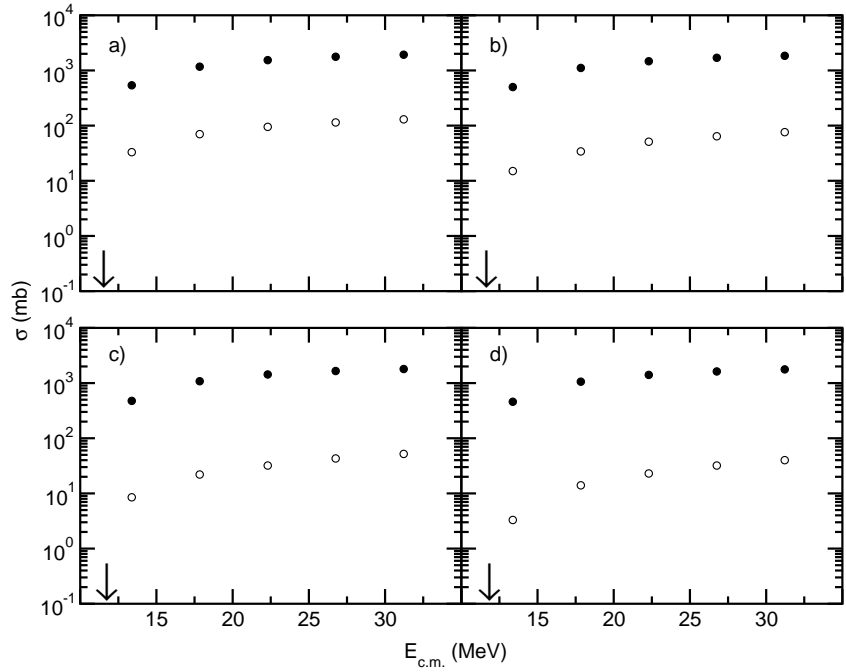}
\end{center}
\caption{Model breakup calculations for a $^{58}$Ni target with
breakup thresholds of: a) 1.0 MeV, b) 1.5 MeV, c) 2.0 MeV and d)
2.5 MeV. The filled and open circles denote the total reaction
cross section and total breakup cross section, respectively. The
arrows indicate the positions of the respective Coulomb barriers,
defined as the maximum of the combined bare nuclear plus Coulomb
potentials. \label{fig:bu58ni}}
\end{figure}
\begin{figure}
\epsfysize=9.0cm
\begin{center}
  \includegraphics[width=.6\textwidth]{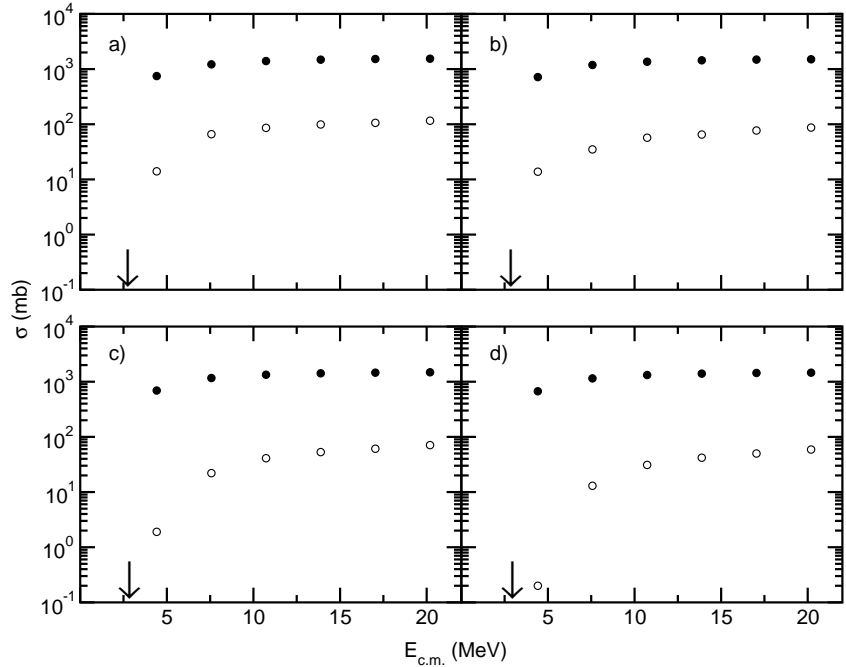}
\end{center}
\caption{Model breakup calculations for a $^{12}$C target with
breakup thresholds of: a) 1.0 MeV, b) 1.5 MeV, c) 2.0 MeV and d)
2.5 MeV. The filled and open circles denote the total reaction
cross section and total breakup cross section, respectively. The
arrows indicate the positions of the respective Coulomb barriers,
defined as the maximum of the combined bare nuclear plus Coulomb
potentials. \label{fig:bu12c}}
\end{figure}

\subsubsection{\it Results of breakup calculations
\label{subsubsec:breakup_results}}

The results are presented in Figs.\
\ref{fig:bu208pb}--\ref{fig:bu12c} as excitation functions of
total reaction cross section and total breakup cross section.

A number of observations are immediately apparent:
\begin{itemize}
\item With the exception of the results for a breakup threshold of
1.0~MeV and a $^{208}$Pb target the total breakup cross section
makes a small or negligible contribution to the total reaction
cross section.

\item The total breakup cross section is very sensitive to the
breakup threshold for the $^{208}$Pb target, less so for $^{58}$Ni
and $^{12}$C, diminishing with increasing threshold energy, as
might be expected.

\item The behaviour as a function of energy of the ratio of total
breakup cross section to total reaction cross section for the
$^{208}$Pb target differs from that for the $^{58}$Ni and $^{12}$C
targets, i.e. for the $^{208}$Pb target this ratio decreases as a
function of increasing incident energy, whereas for the other two
targets the reverse is true.
\end{itemize}

These results show that in terms of total cross section breakup is
a \emph{quantit\'e n\'egligeable}, except for heavy targets at
incident energies close to or below the Coulomb barrier where the
Coulomb force should dominate. In these ``optimal'' conditions the
total breakup cross section is approximately 50\% of the total
reaction cross section for the lowest breakup threshold (1.0~MeV)
at an incident energy of 25~MeV, diminishing rapidly as the breakup
threshold energy increases --- it is of the order of 5\% for a
breakup threshold of 2.5~MeV under the same conditions. However, in
general the magnitude of the cross section is not a reliable guide
to the influence of a coupling on other channels, see e.g.\
\cite{Mac74,Ioa87,Mac87}. Coupling to breakup is known to have an
important effect on the elastic scattering, particularly for
energies near the top of the Coulomb barrier, for a wide range of
targets, see e.g.\ \cite{Sak86}. The first hint of this important
effect came when the hitherto highly successful double-folding
model potential \cite{Sat79} had to be renormalised by a factor of
order 0.5 to fit elastic scattering data for the weakly-bound
nuclei $^6$Li \cite{Sat78}, $^7$Li \cite{Ste80} and $^9$Be
\cite{Sat79a}. This renormalisation was subsequently shown to be
due to a large positive real DPP produced by coupling to breakup
\cite{Sak86,Mac82,Nag82,Hir89}, although the propriety of
simulating what is normally a surface-peaked DPP by a simple
renormalisation of the whole potential is, to say the least,
questionable, particularly if the same potential is to be used to
calculate fusion within the one-dimensional barrier penetration
formalism. However, CC calculations that adopted this procedure to
simulate the effect of couplings to breakup that were not
explicitly included in the coupling scheme proved remarkably
successful in describing fusion data for systems involving weakly
bound projectiles \cite{Ala02}. In fine, the relatively small cross
section belies the importance of breakup in terms of its influence
on other channels.

Before drawing conclusions from our calculations, we shall briefly
show that although they have been considerably simplified they
nevertheless produce qualitatively realistic results. Two recent
coincidence measurements have been performed for the $^6$Li +
$^{208}$Pb \cite{Sig03} and $^6$Li + $^{28}$Si \cite{Pak06} systems
and total $\alpha+d$ breakup cross sections are reported. For the
$^6$Li + $^{208}$Pb system, the total breakup cross sections
\cite{Sig03} at incident energies of 31 and 39~MeV provide
approximately 8\% and 5\%, respectively, of the total reaction
cross section derived from optical model fits \cite{Kee94}. For our
simplified calculations presented in Fig.\ \ref{fig:bu208pb}b this
ratio is approximately 13\% and 7\% at similar incident energies
for a breakup threshold of 1.5~MeV, close to the $\alpha+d$
threshold of $^6$Li. For the $^6$Li + $^{28}$Si system, the ratio
of measured total $\alpha+d$ breakup cross section to total
reaction cross section at an incident energy of 13~MeV is
approximately 2\% \cite{Pak06}. Our simplified calculations for
$^{58}$Ni and $^{12}$C targets and a breakup threshold of 1.5~MeV,
presented in Figs.\ \ref{fig:bu58ni}b and \ref{fig:bu12c}b yield
ratios of approximately 3.5\% and 2\%, respectively, for similar
energies with respect to the Coulomb barrier. When the fact that
the $^6$Li $\rightarrow$ $\alpha+d$ breakup cannot proceed via
dipole excitation is taken into account, and that such excitations
are included in our model calculations which were based on a
``$^7$Li'' projectile, the agreement is rather good. Therefore, our
model calculations are sufficiently realistic to draw reliable
conclusions.

\begin{figure}
\epsfysize=9.0cm
\begin{center}
  \includegraphics[width=.6\textwidth]{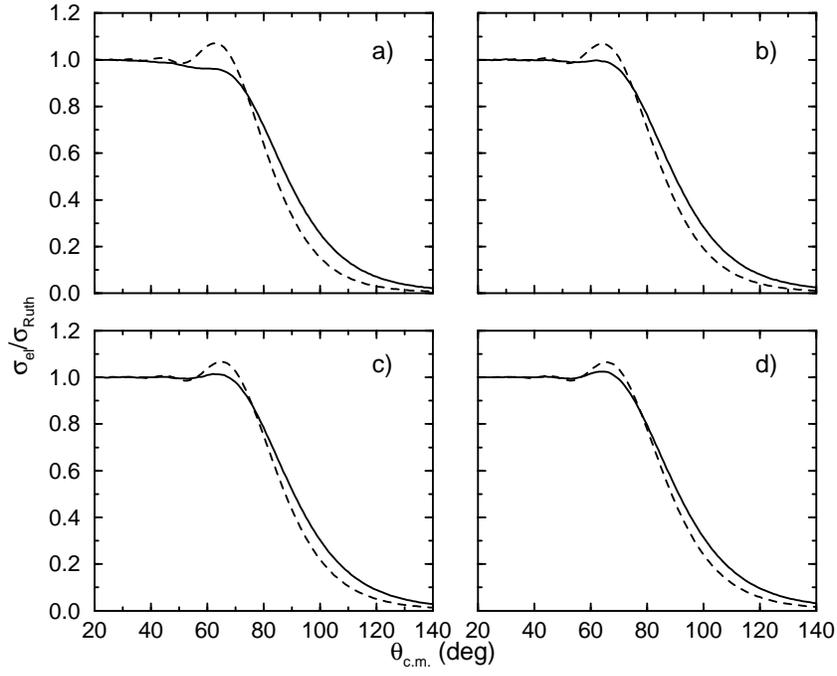}%el208pbtest}
\end{center}
\caption{Model calculations for elastic scattering from a
$^{208}$Pb target at 35~MeV incident energy with breakup thresholds
of a) 1.0 MeV, b) 1.5 MeV, c) 2.0 MeV and d) 2.5 MeV. The solid and
dashed curves denote calculations with and without coupling to
breakup, respectively. \label{fig:el208pb}}
\end{figure}

\begin{figure}
\epsfysize=9.0cm
\begin{center}
  \includegraphics[width=.6\textwidth]{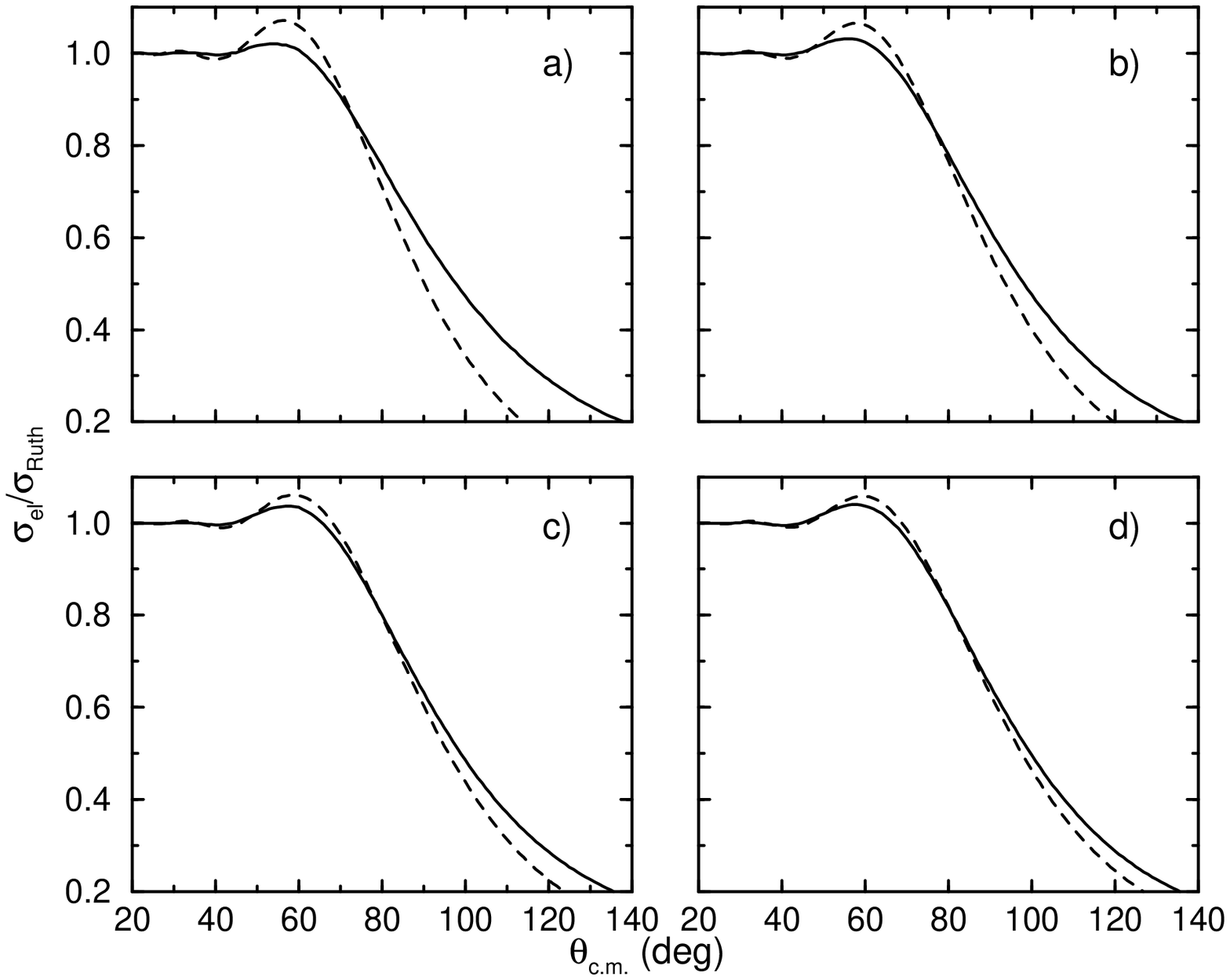}%el58nitest}
\end{center}
\caption{Model calculations for elastic scattering from a $^{58}$Ni
target at 15~MeV incident energy with breakup thresholds of a) 1.0
MeV, b) 1.5 MeV, c) 2.0 MeV and d) 2.5 MeV. The solid and dashed
curves denote calculations with and without coupling to breakup,
respectively. \label{fig:el58ni}}
\end{figure}

\begin{figure}
\epsfysize=9.0cm
\begin{center}
  \includegraphics[width=.6\textwidth]{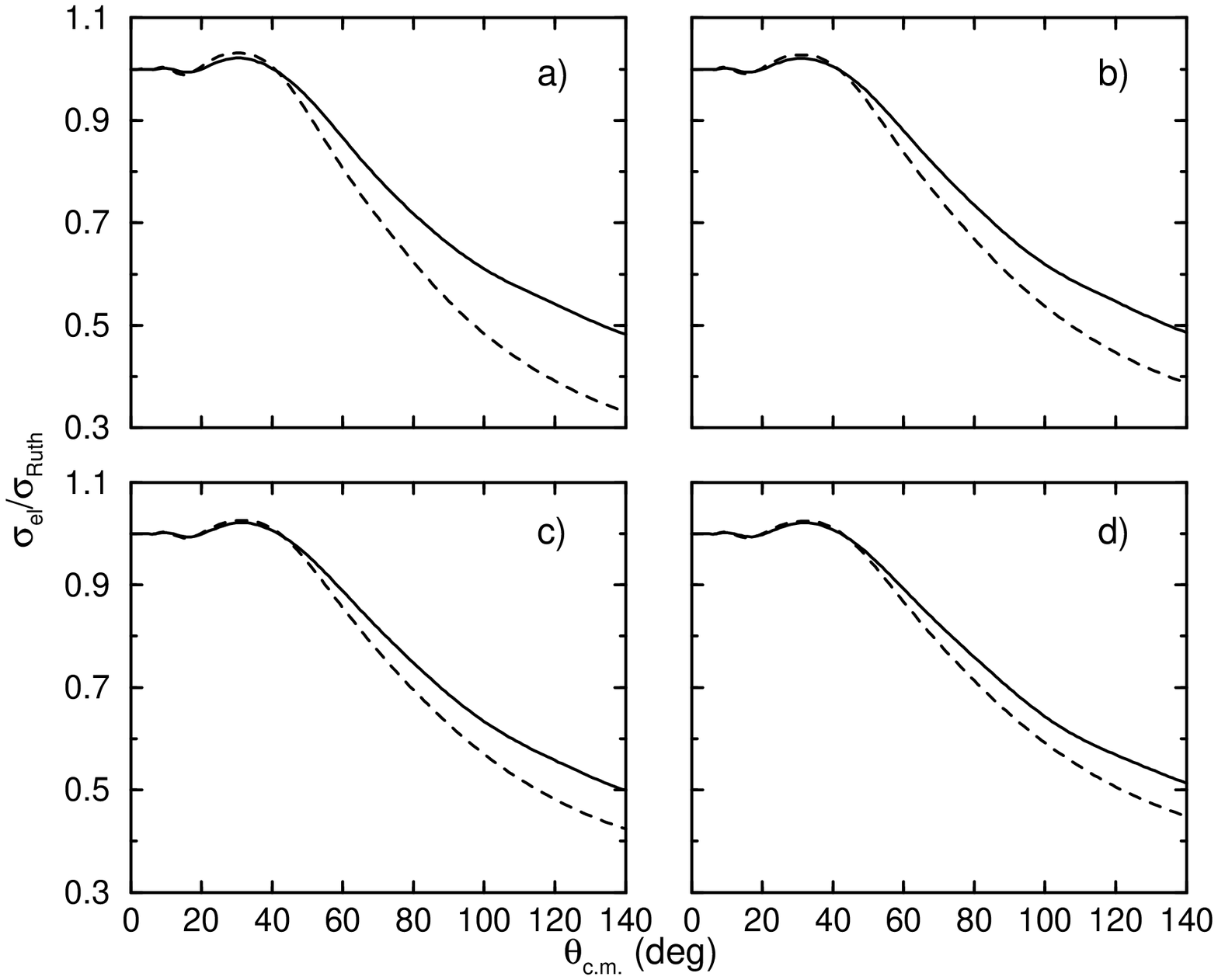}%el12c}
\end{center}
\caption{Model calculations for elastic scattering from a $^{12}$C
target at 5.4~MeV incident energy with breakup thresholds of a) 1.0
MeV, b) 1.5 MeV, c) 2.0 MeV and d) 2.5 MeV. The solid and dashed
curves denote calculations with and without coupling to breakup,
respectively. \label{fig:el12c}}
\end{figure}

Our conclusion is that the total breakup cross section is very
sensitive to both the breakup threshold energy and the target,
i.e.\ the relative importance of Coulomb effects. This sensitivity
is in addition to any differences due to nuclear structure effects,
e.g.\ while the $^6$Li $\rightarrow$ $\alpha+d$ breakup has no
dipole component, the $^6$He $\rightarrow$ $\alpha+2n$ breakup has
a very strong dipole contribution. Therefore, drawing general
inferences as to the effect of breakup on other reaction channels,
in particular elastic scattering and fusion, from data for a single
weakly-bound nucleus and a limited range of targets could be
misleading. This may be particularly so when results obtained for
the stable weakly-bound nuclei, $^6$Li, $^7$Li and $^9$Be, are used
to infer the influence of breakup for the weakly-bound halo nuclei
such as $^6$He and $^{11}$Li. This is particularly striking when we
compare the elastic scattering angular distributions for the
different breakup thresholds and targets obtained from our model
calculations. In Figs.\ \ref{fig:el208pb}--\ref{fig:el12c} we plot
the elastic scattering angular distributions at incident energies
of 35~MeV, 15~MeV and 5.4~MeV for the $^{208}$Pb, $^{58}$Ni and
$^{12}$C targets, respectively. These energies yield
$E_{\mathrm{c.m.}}/V_{\mathrm{B}}$ values of approximately 1.15,
1.18 and 1.21, respectively, chosen as coupling effects are usually
largest for incident energies around 10\% to 20\% above the Coulomb
barrier.


It will be noted that the elastic scattering angular distributions
for a breakup threshold of 1.0~MeV are clearly distinguishable from
those for the higher thresholds, and that the angular range over
which the main difference between the results of calculations with
and without coupling is apparent moves away from the Coulomb
rainbow region to backward angles in passing from the $^{208}$Pb to
the $^{12}$C target. The breakup coupling effect is also energy
dependent; for the $^{12}$C target at an incident energy of 12~MeV
the angular distributions for the different breakup thresholds are
indistinguishable. For a breakup threshold of 1.0~MeV the effect is
much more dramatic for the $^{208}$Pb target, at least in the
angular region of the Coulomb rainbow, due to the effect of Coulomb
breakup. This is in contrast to higher breakup thresholds where the
influence of Coulomb breakup couplings on the elastic scattering
has been found to be small, even for incident energies close to the
Coulomb barrier, despite its importance for the total breakup cross
section \cite{Kee96}.

We thus see that breakup is a complex process, and that each system
needs to be investigated on its own merits if we are to draw
reliable conclusions concerning its influence on other reaction
channels. This is particularly so in the case of halo nuclei, as
their very low breakup thresholds lead to much stronger effects.
However, even for halo nuclei, breakup is not an important
contributor to the total reaction cross section except for systems
where the Coulomb interaction dominates and then only for incident
energies close to or below the top of the Coulomb barrier. It is
therefore possible that the apparent contradictions as to whether
sub-barrier fusion enhancement has been observed or not in
different systems involving weakly-bound nuclei may be partly due
to this ``system dependence'' --- a system involving fusion with a
lighter target may still exhibit an enhancement, but the effect may
be too small to be observed. However, as we shall see in the next
section, there are other system dependent effects that may be
stronger than those due to breakup which further complicate the
matter.

\subsection{\it Transfer reactions and their effect on other reaction
channels \label{subsec:transfer}}

As stated in section \ref{subsec:methods}, in contrast to breakup,
we have a reliable means of calculating the influence of transfer
reactions on the fusion cross section through CRC calculations
employing the ingoing-wave boundary condition or its equivalent. In
this section we present such calculations for
$^{208}$Pb($^6$He,$^5$He)$^{209}$Pb,
$^{208}$Pb($^8$He,$^7$He)$^{209}$Pb,
$^{208}$Pb($^6$Li,$^5$Li)$^{209}$Pb and
$^{60}$Ni($^6$Li,$^5$Li)$^{61}$Ni as representative examples. All
calculations were carried out using the code \textsc{fresco}
\cite{Tho88}. That transfer couplings can have significant effects
on the elastic scattering for weakly bound nuclei was shown in the
case of the $^9$Be + $^{208}$Pb system \cite{Kee05}. It was found
that couplings to the positive $Q$-value
$^{208}$Pb($^9$Be,$^8$Be)$^{209}$Pb transfer reaction produced a
DPP with the same characteristics in the surface region as that
produced by breakup, leading to a reduction in the real part of the
effective potential and thus an increase in the effective Coulomb
barrier. Thus, couplings to transfer reactions may be responsible
for part of the reduction in the double folding model potentials
found necessary for weakly bound nuclei referred to in section
\ref{subsec:breakup}, and it is reasonable to suppose that they may
also have important effects on fusion induced by such nuclei.

\subsubsection{\it Conditions for the transfer calculations
\label{subsubsec:tr_ansatz}}

We adopted a similar method to that used in \cite{Tho89}, whereby
our ``bare'' potentials in all channels consisted of a
double-folded real part and an interior Woods-Saxon imaginary part
to simulate the ingoing-wave boundary condition \cite{Rho84}. We
used the M3Y effective interaction in the form of \cite{Sat79} and
took the parameterisations of \cite{Tan92} for the $^6$He and
$^8$He matter densities, the $^6$Li matter density being taken from
\cite{Coo83}. For the $^5$He and $^7$He matter densities we took
the same form as \cite{Tan92}, adjusting the parameters of the
neutron part to give matter radii of 1.85 fm and 2.65 fm,
respectively, close to the calculated values of \cite{Ste97} and
\cite{Nef04}. We assumed the mirror hypothesis for $^5$Li, taking
the proton density as equal to the $^5$He neutron density and
\emph{vice versa}, as sufficiently realistic for our purposes. The
$^{208}$Pb matter density was derived from the charge density of
\cite{Eut78}, unfolding the finite charge distributions of the
proton and neutron as in \cite{Sat79} and assuming that
$\rho_{\mathrm{n}} = (N/Z) \rho_{\mathrm{p}}$. The $^{209}$Pb
density was taken to be the same as that for $^{208}$Pb. The
$^{60}$Ni matter density was taken from \cite{Lom81}, the $^{61}$Ni
density being taken to be equal to that for $^{60}$Ni. The
Woods-Saxon imaginary potentials had parameters: $W = 50$~MeV, $R_W
= 1.0 \times (\mathrm{A_p}^{1/3} + \mathrm{A_t}^{1/3})$~fm, $a_W =
0.3$~fm. The double-folded potentials were calculated with the code
\textsc{DFPOT}~\cite{Coo82}.

Couplings for single neutron transfer to the 0.0~MeV $9/2^+$,
1.57~MeV $5/2^+$, 2.03~MeV $1/2^+$, 2.49~MeV $7/2^+$ and 2.54~MeV
$3/2^+$ states of $^{209}$Pb were included in the
$^{208}$Pb($^6$He,$^5$He), $^{208}$Pb($^8$He,$^7$He) and
$^{208}$Pb($^6$Li,$^5$Li) calculations. The $n$ + $^{208}$Pb
binding potentials and spectroscopic factors were taken from the
adiabatic model $^{208}$Pb($d$,$p$) analysis of \cite{Kov74}. The
$n$ + $^5$He, $n$ + $^7$He and $n$ + $^5$Li binding potentials were
of Woods-Saxon form, with $R_0 = 1.25 \times \mathrm{A}^{1/3}$ fm,
$a = 0.65$ fm and a spin-orbit term with the same geometry and a
depth of 6~MeV, the depth of the central part being adjusted to
obtain the correct binding energy. The $n$ + $^5$He, $n$ + $^7$He
and $n$ + $^5$Li spectroscopic factors were taken from
\cite{Nem88}, \cite{Ska05} and \cite{Coh67}, respectively.
Stripping to the 1.27~MeV and 1.49~MeV $1/2^-$ states of $^5$He and
$^5$Li, respectively, was included in addition to that to the
$3/2^-$ ground state resonances. For the $^{60}$Ni($^6$Li,$^5$Li)
calculations transfers to the 0.0~MeV $3/2^-$, 0.067~MeV $5/2^-$,
0.283~MeV $1/2^-$, 0.909~MeV $5/2^-$, 1.132~MeV $5/2^-$, 1.185~MeV
$3/2^-$, 2.122~MeV $9/2^+$, 2.123~MeV $1/2^-$ and 3.482~MeV $9/2^+$
states of $^{61}$Ni were included, the $n$ + $^{60}$Ni binding
potentials and spectroscopic factors being taken from \cite{Wes91}.

\subsubsection{\it Validity of transfer calculations: comparison to
experimental data
\label{subsubsec:tr_comp}}

\begin{figure}
\epsfysize=9.0cm
\begin{center}
  \includegraphics[width=.5\textwidth]{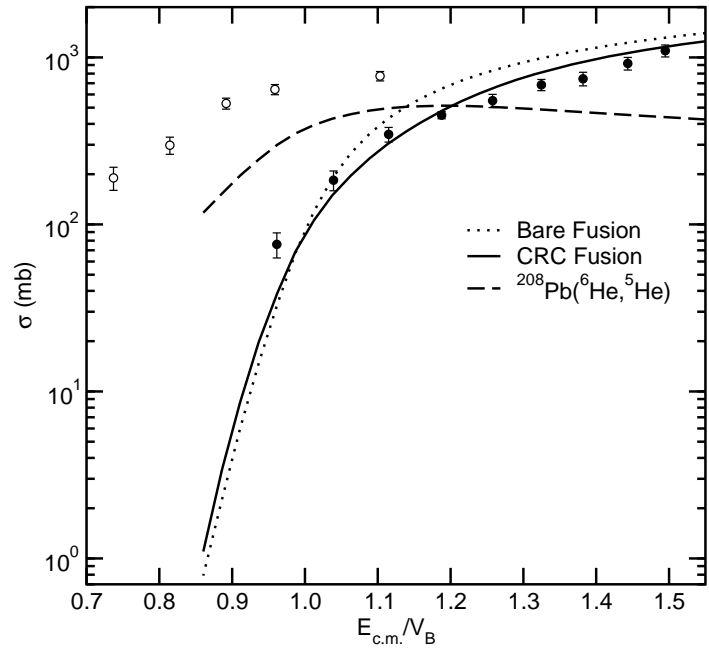}
\end{center}
\caption{Calculated $^6$He + $^{208}$Pb total fusion and
$^{208}$Pb($^6$He,$^5$He)$^{209}$Pb excitation functions. The
filled and open circles denote the total fusion and total $\alpha$
production cross sections for the $^6$He + $^{209}$Bi system of
\cite{Kol98} and \cite{Agu01}, respectively.
\label{fig:fus6he5he}}
\end{figure}
\begin{figure}
\epsfysize=9.0cm
\begin{center}
  \includegraphics[width=.5\textwidth]{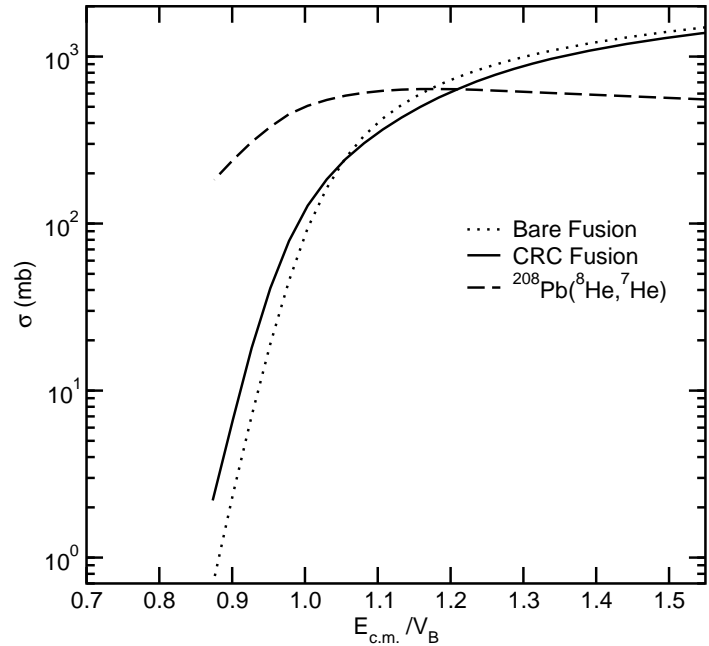}
\end{center}
\caption{Calculated $^8$He + $^{208}$Pb total fusion and
$^{208}$Pb($^8$He,$^7$He)$^{209}$Pb excitation functions.
\label{fig:fus8he7he}}
\end{figure}
\begin{figure}
\epsfysize=9.0cm
\begin{center}
  \includegraphics[width=.5\textwidth]{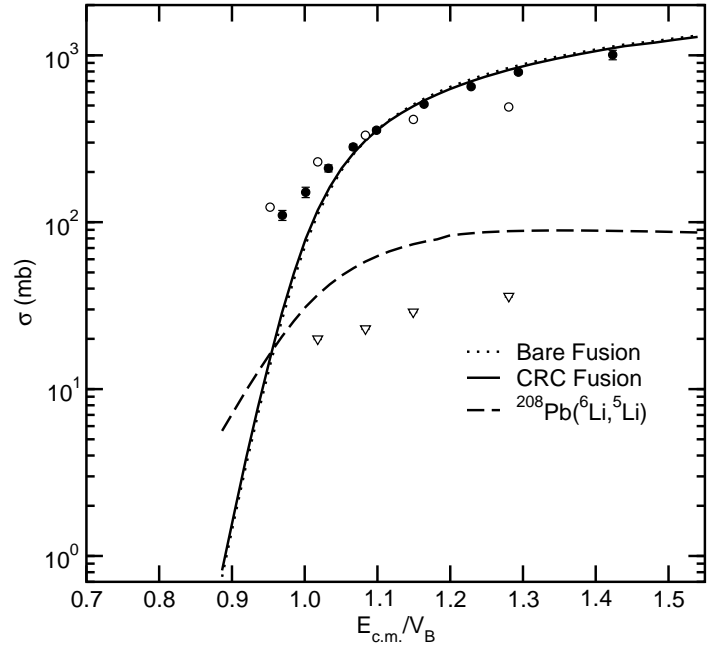}
\end{center}
\caption{Calculated $^6$Li + $^{208}$Pb total fusion and
$^{208}$Pb($^6$Li,$^5$Li)$^{209}$Pb excitation functions. The
filled circles denote the total fusion cross sections for the
$^6$Li + $^{209}$Bi system \cite{Das04}, while the open circles
and triangles denote the total $\alpha$ production and total
$\alpha$ + $p$ coincidence cross sections, respectively, for the
$^6$Li + $^{208}$Pb system \cite{Sig03}. \label{fig:fus6li5li}}
\end{figure}
\begin{figure}
\epsfysize=9.0cm
\begin{center}
  \includegraphics[width=.5\textwidth]{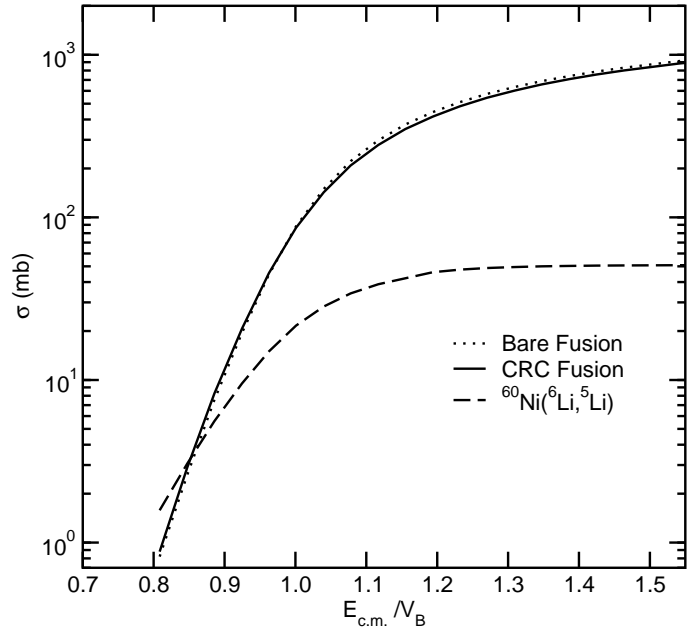}
\end{center}
\caption{Calculated $^6$Li + $^{60}$Ni total fusion and
$^{60}$Ni($^6$Li,$^5$Li)$^{61}$Ni excitation functions.
\label{fig:fusni6li5li}}
\end{figure}

Calculated excitation functions for the total one neutron transfer
reaction, ``bare'' (i.e.\ no coupling) and CRC fusion cross
sections are presented in Figs.\
\ref{fig:fus6he5he}--\ref{fig:fusni6li5li} as functions of
$E_{\mathrm{c.m.}}/V_{\mathrm{B}}$. The Coulomb barriers were
calculated as the maxima of the combined ``bare'' nuclear plus
Coulomb potentials in the respective entrance channels. We also
plot data for the complete fusion of $^6$He \cite{Kol98} and $^6$Li
\cite{Das04} with $^{209}$Bi and the total $\alpha$ production
cross sections for the $^6$He + $^{209}$Bi \cite{Agu01} and $^6$Li
+ $^{208}$Pb \cite{Sig03} systems, plus the total $\alpha+p$
coincidence cross sections for $^6$Li + $^{208}$Pb \cite{Sig03}.
The total fusion cross sections for $^6$He and $^6$Li + $^{209}$Bi
should be similar to those for a $^{208}$Pb target. The total
$\alpha$ production cross section for $^6$He + $^{209}$Bi should
provide a reasonable estimate of that for a $^{208}$Pb target,
although nuclear structure differences may be more important here.
We have chosen $^{208}$Pb in preference to $^{209}$Bi in order to
have tractable calculations.

Before considering the effect on fusion of these couplings we shall
compare the calculated transfer cross sections with existing
measurements. For the $^6$He + $^{209}$Bi system, $n+\alpha$
coincidence measurements indicate a total one neutron transfer
cross section of $(155 \pm 25)$~mb at an incident energy of
22.9~MeV \cite{Byc04,DeY05}, accounting for $(20 \pm 2)$\% of the
total $\alpha$ production cross section. Our calculated value at
this energy is 510~mb, about three times larger. It is possible
that part of the discrepancy is due to nuclear structure
differences between $^{208}$Pb and $^{209}$Bi --- $^{208}$Pb is a
very good core for a single-particle description of the
\nuc{209}{Pb} nucleus, thus making calculation of the transfer much
easier --- but it is most likely due to the effect of other
couplings not included in the calculations, such as breakup or two
neutron transfer. Unfortunately, a meaningful calculation for the
latter process is not possible, even for a $^{208}$Pb target, due
to lack of knowledge of the structure of the excited states of
$^{210}$Pb in the most important excitation energy region
--- even spins and parities are unknown. The situation is much
worse for $^{211}$Bi if one contemplated a similar calculation for
a $^{209}$Bi target, as not only are the spins and parities of the
relevant states unknown, their number is considerably greater than
in $^{210}$Pb. The cross section for two neutron transfer is found
to be much larger than that for single neutron transfer in the
$^6$He + $^{209}$Bi system, approximately $(400 \pm 100)$~mb at
angles greater than or equal to that of the ``grazing peak'', or
$(55 \pm 12)$\% of the total $\alpha$ production cross section
\cite{DeY05}. Thus it is possible that two-step transfer via the
($^6$He,$^5$He), ($^5$He,$^4$He) process may reduce the one neutron
transfer cross section, although it is unlikely to account for the
whole of the discrepancy. We note that test calculations using a
$^5$He + $^{209}$Pb double-folding potential calculated using a
$^5$He density with a larger r.m.s.\ matter radius did lead to a
slight reduction in the calculated one neutron transfer cross
section, although only of the order of a few percent. However,
tests also found that including surface absorption, i.e.\
increasing the radius and diffuseness parameters of the imaginary
potentials to more conventional values, considerably reduced the
calculated transfer cross sections to better match the measured
value. As increased surface absorption may be considered to
simulate the effect of couplings to other channels not explicitly
included, particularly the two neutron transfer which should
contribute a strong surface absorption due to its large cross
section, we conclude that the present calculations are adequate to
investigate the effect of a single process, i.e. the
($^6$He,$^5$He) single neutron transfer on the total fusion cross
section.
\begin{figure}
\epsfysize=9.0cm
\begin{center}
  \includegraphics[width=.3\textwidth]{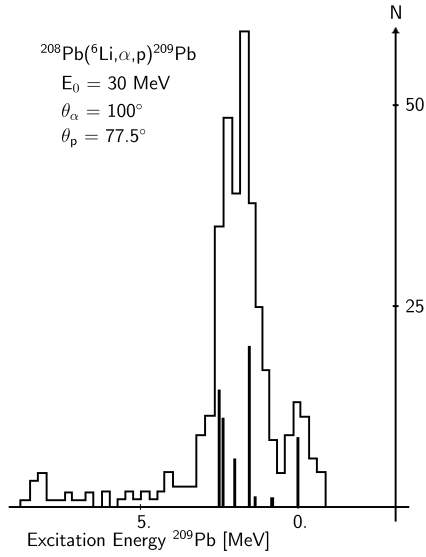}
\end{center}
\caption{Projected $^{208}$Pb($^6$Li, $\alpha$, $p$)$^{209}$Pb
spectrum adapted from \cite{Ost74}. The vertical bars indicate the
total cross sections for the $^{208}$Pb($d$, $p$)$^{209}$Pb
reaction. \label{fig:spec6li5li}}
\end{figure}

Our calculated $^{208}$Pb($^6$Li, $^5$Li)$^{209}$Pb cross sections
are approximately a factor of 2.5 greater than the measured total
$\alpha + p$ cross sections of \cite{Sig03} and account for about
18\% of the measured total $\alpha$ production cross section
\cite{Sig03}. Single neutron transfer was not considered as a
possible production mechanism for the observed $\alpha + p$
coincidences in \cite{Sig03}, although it was discussed in an
earlier article by the same group \cite{Sig01}. However, $\alpha +
p$ coincidences in the $^6$Li + $^{208}$Pb system have been
previously observed in this energy range \cite{Ost74} and the
projected spectrum shown in Fig.\ \ref{fig:spec6li5li} clearly
shows that this is indeed a transfer process populating known bound
states in $^{209}$Pb. The reaction was found to proceed mainly via
the ground state of $^5$Li \cite{Ost74} and our calculations are
consistent with this, the total transfer cross section with $^5$Li
in its ground state is approximately a factor of 9 greater than
that with $^5$Li in its first excited state at a $^6$Li incident
energy of 30~MeV. Similar remarks concerning the effect of
increased surface absorption also apply here.

\begin{figure}
\epsfysize=9.0cm
\begin{center}
  \includegraphics[width=.5\textwidth]{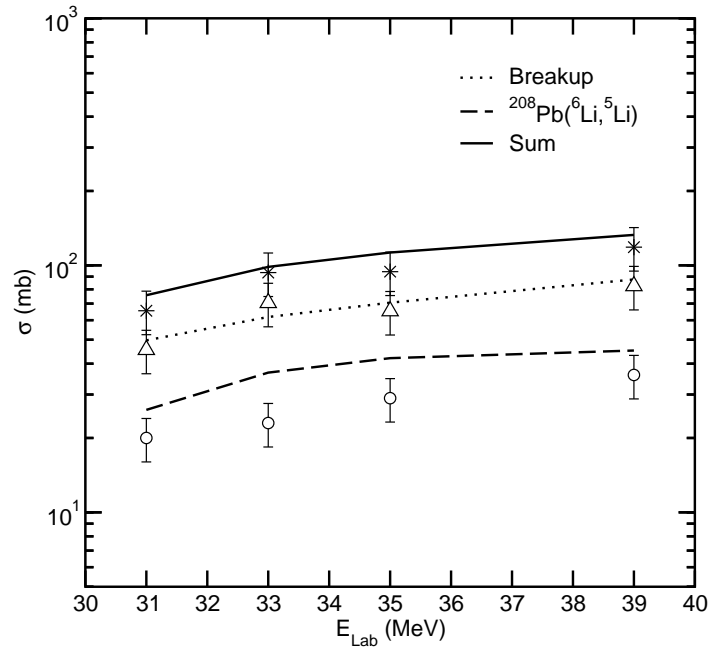}
\end{center}
\caption{Results of more refined calculations of breakup and
transfer for the $^6$Li + $^{208}$Pb system. The data are taken
from \cite{Sig03} and represent the $\alpha+p$ coincidences
(circles), $\alpha+d$ coincidences (triangles) and their sum
(stars). The $\alpha+p$ coincidences are modelled as a
$^{208}$Pb($^6$Li, $^5$Li)$^{209}$Pb transfer process.
\label{fig:buxsec}}
\end{figure}

More refined calculations intended to model the direct reaction
processes as accurately as possible are able to provide a good
description of the available elastic scattering and $\alpha+d$ and
$\alpha+p$ coincidence data for the $^6$Li + $^{208}$Pb system. The
$^6$Li $\rightarrow$ $\alpha+d$ breakup was included using the CDCC
technique as in \cite{Kee03} and the $^{208}$Pb($^6$Li,
$^5$Li)$^{209}$Pb transfer was included as described above, with
the exception that the $^5$Li + $^{209}$Pb potential was calculated
using the global $^6$Li parameters of \cite{Coo82a} --- tests found
that the results are not very sensitive to the choice of exit
channel potential. Results are compared with the data of
\cite{Sig03} in Fig.\ \ref{fig:buxsec}.
Agreement between calculations and data is good (we have added
error bars of 20\% to the data, indicated in \cite{Sig03} as being
the absolute accuracy of the exclusive data) with the total
calculated $^{208}$Pb($^6$Li, $^5$Li)$^{209}$Pb cross sections
approximately 20\% larger than the $\alpha+p$ coincidence data, a
discrepancy easily accounted for by uncertainties in the $^6$Li
$\rightarrow$ $^5$Li + $n$ form factor.

\subsubsection{\it Results of calculations: effect of the
couplings on fusion
 \label{subsubsec:tr_results}}

Turning now to the effect of the transfer couplings on the
calculated total fusion cross sections, we immediately see from
Figs.\ \ref{fig:fus6he5he}-\ref{fig:fusni6li5li} that qualitatively
all the systems studied show the same behaviour, i.e.\ reduction
above the barrier changing to enhancement below it. However,
quantitatively, the two halo nuclei give very different results to
those for $^6$Li. Whereas the coupling effects are significant for
$^6$He and $^8$He, both above and below the barrier, they are
completely negligible for $^6$Li for both targets. This is not
simply a $Q$-value effect, as Table \ref{tab:qval} shows ---
$^{60}$Ni was deliberately chosen to give a ($^6$Li,$^5$Li)
$Q$-value close to that for $^{208}$Pb($^6$He,$^5$He)$^{209}$Pb ---
although this does play a r\^ole.
\begin{table}
 \renewcommand{\arraystretch}{1.2}
 \begin{center}
  \caption{Reaction $Q$-values for the transfers considered in this
  section.
  \label{tab:qval}}\bigskip
  \begin{tabular}{l c}
   \hline \hline \\
   [-5mm] Reaction & $Q$-value \\
   \hline \\
   [-4mm]
   $^{208}$Pb($^6$He,$^5$He)$^{209}$Pb & $+2.07$ MeV \\
   $^{208}$Pb($^8$He,$^7$He)$^{209}$Pb & $+1.35$ MeV \\
   $^{208}$Pb($^6$Li,$^5$Li)$^{209}$Pb & $-1.73$ MeV \\
   $^{60}$Ni($^6$Li,$^5$Li)$^{61}$Ni & $+2.16$ MeV \\
   \hline \hline
  \end{tabular}
 \end{center}
\end{table}
The largest effect is due to the magnitude of the spectroscopic
amplitudes (essentially $\sqrt{C^2S}$) --- those for the
$^6$He/$^5$He overlap being approximately twice those for
$^6$Li/$^5$Li and that for $^8$He/$^7$He nearly three times as
large. Test calculations for $^{208}$Pb($^6$Li,$^5$Li)$^{209}$Pb
with the $^6$Li/$^5$Li spectroscopic amplitudes doubled showed a
considerable increase in the effect on the calculated fusion cross
section, in both the increase for sub-barrier energies and the
decrease above the barrier.

The observation of this effect, i.e.\ sub-barrier enhancement and
above-barrier reduction of the fusion cross section for coupling to
reaction channels with positive $Q$-values, is not new, as the
schematic two-channel calculations of \cite{Das83} and \cite{Tho89}
yield similar results, although the more sophisticated calculations
of \cite{Tho89} yield a much smaller above barrier reduction of the
fusion cross section for a positive $Q$-value when the couplings
are treated as transfers rather than inelastic excitations. We also
note that similar coupling effects on the calculated fusion cross
section to those seen here were observed in an extensive analysis
of  the $^{18}$O + $^{60}$Ni system \cite{Per06}.

A comparison of the calculated total fusion cross sections with the
measured values for the $^6$He + $^{209}$Bi and $^6$Li + $^{209}$Bi
systems in Figs.\ \ref{fig:fus6he5he} and \ref{fig:fus6li5li} shows
that at energies above the barrier our calculations provide a
rather good description of the data. However, there is an important
difference between the two systems in that the effect of the
$^{208}$Pb($^6$He,$^5$He)$^{209}$Pb transfer couplings is essential
to the good agreement between calculations and data, whereas for
the $^6$Li + $^{209}$Bi system the single neutron transfer
couplings have a negligible effect. At sub-barrier energies the
$^6$Li + $^{209}$Bi system shows a clear enhancement of the total
fusion cross section with respect to both the ``bare'' no coupling
and the CRC calculations, while for the $^6$He + $^{209}$Bi system
an enhancement is only seen at the lowest measured energy. For the
$^6$Li + $^{209}$Bi system the transfer couplings have a negligible
effect while for the $^6$He + $^{209}$Bi system the effect is too
small to give good agreement with the lowest energy data point.
Thus, we see that there are evidently other important couplings
that must be included in the $^6$Li + $^{209}$Bi system, and
possibly in the $^6$He + $^{209}$Bi system, if we are to obtain a
good description of the total fusion excitation function over the
whole energy range. These couplings could be breakup, target
inelastic excitations or other transfer reactions, particularly the
($^6$He,$^4$He) two-neutron transfer for the $^6$He + $^{209}$Bi
system.

\begin{figure}
\epsfysize=9.0cm
\begin{center}
  \includegraphics[width=.5\textwidth]{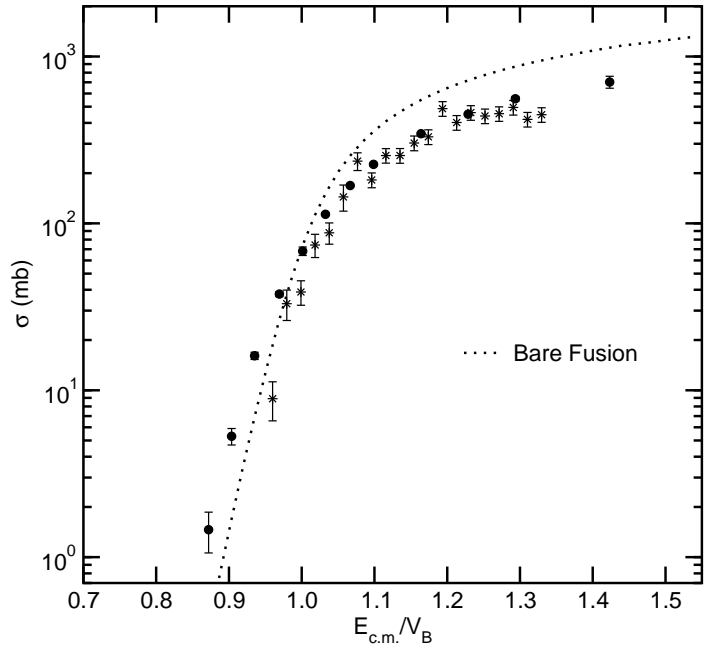}
\end{center}
\caption{Calculated ``bare'' $^6$Li + $^{208}$Pb fusion compared
to measured $^6$Li + $^{209}$Bi \cite{Das04} (filled circles) and
$^6$Li + $^{208}$Pb \cite{Wu03} (stars) {\em complete} fusion
excitation functions. \label{fig:comp6li208pb}}
\end{figure}
We finally consider the effects of couplings on the \emph{complete}
fusion cross section. In Fig.\ \ref{fig:comp6li208pb} we compare
the bare fusion calculation for $^6$Li + $^{208}$Pb to data for
complete fusion in the $^6$Li + $^{209}$Bi \cite{Das04} and $^6$Li
+ $^{208}$Pb \cite{Wu03} systems. The data for the two systems
agree well for energies above the Coulomb barrier while at
sub-barrier energies the $^6$Li + $^{208}$Pb cross section falls
more rapidly with energy. It is evident that a substantial fusion
suppression is needed at above barrier energies in order to match
either data set, while at sub-barrier energies the $^6$Li +
$^{209}$Bi data exhibit enhancement whereas the $^6$Li + $^{208}$Pb
data show a slight suppression. If this is a real effect it is
probably not due to breakup, as the target is essentially a
spectator in the breakup process and the extra unit of charge on
the $^{209}$Bi target would not be expected to be important.

To summarise, we find that transfer couplings, although rather
neglected in the debate concerning the possible enhancement of
fusion for halo nuclei, can have important effects. We also note
that these effects are not limited to the sub-barrier region but
are also important at energies above the barrier. In this respect
our conclusions are similar to those of a detailed study of the
$^{18}$O + $^{60}$Ni system \cite{Per06}. It is again evident that
attempting to draw conclusions concerning halo nuclei from studies
of the stable weakly-bound nuclei could be misleading.

%\newpage

\section{Discussion
\label{sec:discussion}}

We have seen in the previous section that, in contrast to the
stable weakly-bound nucleus $^6$Li, couplings to single neutron
transfer reactions have an important influence on fusion induced by
the neutron halo nuclei $^6$He and $^8$He. This influence is
manifest as an above-barrier reduction and a sub-barrier
enhancement of the calculated fusion cross section relative to the
no-coupling case, the above-barrier reduction being much more
important than the sub-barrier enhancement for $^6$He, less so for
$^8$He. While similar couplings produce a qualitatively similar
effect for $^6$Li, it is entirely negligible. In this section we
shall attempt to establish whether this is a universal phenomenon
for halo nuclei. We will only briefly discuss the complex and
sometimes controversial question of the effect on fusion of
coupling to breakup, due to the lack of a completely adequate
model. The conclusions of the original analyses of the fusion data
for weakly bound exotic nuclei presented in section
\ref{subsec:exp_results} fall into two groups, those that infer a
sub-barrier enhancement of the fusion cross section: $^6$He +
$^{206}$Pb \cite{Pen06} and $^6$He + $^{209}$Bi \cite{Kol98}, and
those that infer no enhancement: $^6$He + $^{64}$Zn \cite{DiP04},
$^6$He + $^{238}$U \cite{Raa04}, $^7$Be + $^{238}$U \cite{Raa06}
and $^{11}$Be + $^{209}$Bi \cite{Sig98}. However, enhancement (or
lack of it) is inferred with respect to different references and
different methods of ``reducing'' the data are adopted in the
individual articles. In this section we shall also investigate
whether a consistent analysis method can fit the ensemble of these
data into a single, coherent picture.

\subsection{\it Comparison between experimental data and
calculations: conditions
\label{subsec:disc_conditions}}

In order to motivate the discussion, we compare the data for fusion
induced by weakly bound radioactive nuclei on various targets
already presented in section~\ref{subsec:exp_results} with
one-dimensional barrier penetration model calculations employing
double-folded potentials. We also present similar calculations for
fusion induced by the core nuclei interacting with the same
targets, where available, for comparison purposes. The calculations
are similar to those already described in section
\ref{subsec:transfer}; we employ the same standard M3Y interaction
with the $^6$He matter density taken from \cite{Tan92}, the $^7$Be
matter density from \cite{Bha00}, the $^{10}$Be and $^{11}$Be
matter densities from \cite{Sag92} and the $^4$He matter density
derived from the three parameter Fermi distribution charge density
of \cite{Mcc77} by unfolding the proton and neutron charge
distributions as in \cite{Sat79} and assuming that
$\rho_{\mathrm{n}} = (N/Z) \rho_{\mathrm{p}}$. Target matter
densities were derived from charge densities in a similar way to
that for $^4$He, the references being as follows: $^{64}$Zn
\cite{Keg77}, $^{209}$Bi \cite{Eut78} and $^{238}$U \cite{Coo76}.
The proton density for $^{206}$Pb was obtained by unfolding the
finite neutron and proton charge distributions from the charge
density of \cite{Eut78} while the neutron density was taken from
\cite{Gil76}.

As discussed previously, this method of comparison is not
completely unambiguous, as the double-folded potentials depend on
the choice of effective nucleon-nucleon interaction and matter
densities used to calculate them. We have chosen a standard M3Y
effective interaction with no density dependence for the sake of
simplicity; at the relatively low energies considered here this
should be adequate. Modern density-dependent versions of the M3Y,
e.g.\ the CDM3Y6 \cite{Kho97}, could be used if a more
sophisticated interaction is desired. The procedure used for
extracting the target matter densities from empirical charge
densities, while relatively standard, may be questioned
\cite{Sat79} for nuclei where N $\neq$ Z, even approximately. The
matter densities for the radioactive beam nuclei are of necessity
taken from calculations, although for $^6$He we used a matter
density deduced \cite{Tan92} from interaction cross section
measurements assuming harmonic oscillator forms and yielding
results similar to relativistic mean field calculations.

\begin{figure}
\epsfysize=9.0cm
\begin{center}
  \includegraphics[width=.5\textwidth]{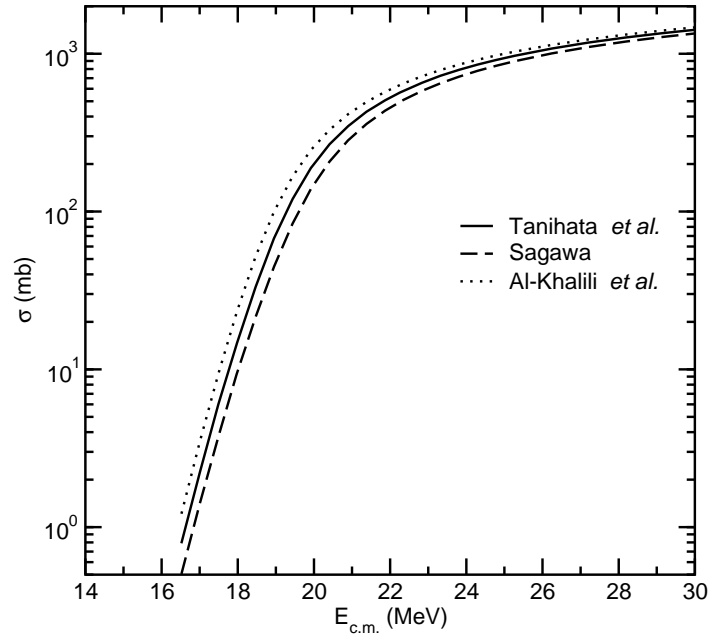}
\end{center}
\caption{Calculated $^6$He + $^{208}$Pb fusion excitation
functions employing the densities of Tanihata {\em et al.\/}
\cite{Tan92} (full curve), Sagawa \cite{Sag92} (dashed curve) and
FC6 of Al-Khalili {\em et al.\/} \cite{Alk96} (dotted
curve).\label{fig:projcomp}}
\end{figure}

\begin{figure}
\epsfysize=9.0cm
\begin{center}
  \includegraphics[width=.5\textwidth]{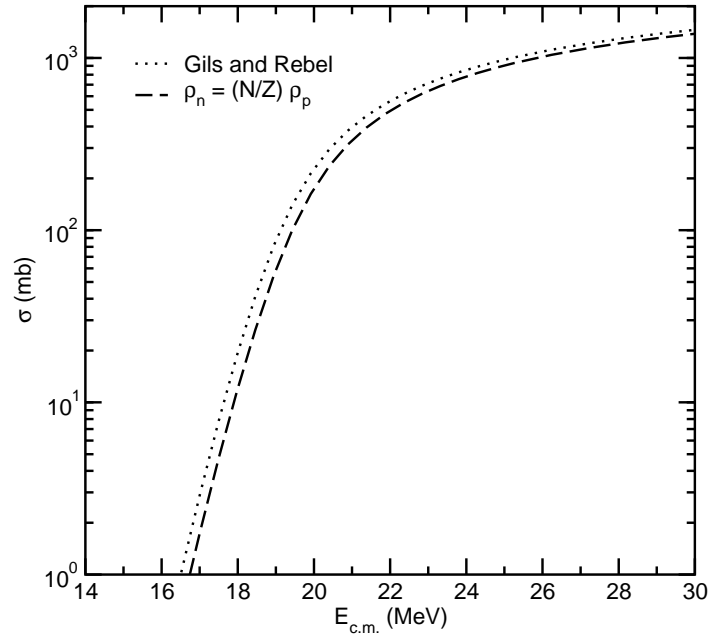}
\end{center}
\caption{Calculated $^6$He + $^{206}$Pb fusion excitation functions
employing different neutron density distributions for $^{206}$Pb,
from Gils and Rebel \cite{Gil76} (dotted curve) and derived from
the proton distribution assuming that $\rho_{\mathrm{n}} = (N/Z)
\rho_{\mathrm{p}}$ (dashed curve).\label{fig:targcomp}}
\end{figure}

To investigate the level of uncertainty in the calculated fusion
excitation functions we present in Figs.\ \ref{fig:projcomp} and
\ref{fig:targcomp} calculations for the $^6$He + $^{208}$Pb and
$^6$He + $^{206}$Pb systems, respectively, where the effects of
choosing different $^6$He matter densities and $^{206}$Pb neutron
densities are tested. Fig. \ref{fig:projcomp} compares three
calculations for the $^6$He + $^{208}$Pb system using different
$^6$He matter densities \cite{Tan92,Sag92,Alk96}.
The excitation functions calculated with the $^6$He densities of
Sagawa \cite{Sag92} and Al-Khalili {\em et al.\/} \cite{Alk96}
show shifts of $\sim +0.3$~MeV and $\sim -0.3$~MeV, respectively,
with respect to that calculated with the density of Tanihata {\em
et al.\/} \cite{Tan92}, consistent with the differences in the
calculated Coulomb barrier heights.
In Fig.\ \ref{fig:targcomp} we compare fusion excitation functions
for the $^6$He + $^{206}$Pb system calculated using two different
$^{206}$Pb neutron density distributions. Both calculations use
$^{206}$Pb proton density distributions derived by unfolding the
finite nucleon charge distributions from the charge density of
\cite{Eut78}. The dotted curve employs the neutron density
distribution of \cite{Gil76} while the dashed curve assumes that
$\rho_{\mathrm{n}} = (N/Z) \rho_{\mathrm{p}}$. Again, we see a
shift of $\sim 0.3$~MeV between the two curves, commensurate with
the shift in the calculated Coulomb barrier heights. We thus see
that the overall level of uncertainty is a shift in energy of the
order of 1~MeV in the calculated fusion excitation function.
Calculations based on these double-folded potentials should
therefore provide a reasonable basis for discussion, although the
details of their comparison with experiment should be treated with
due caution.

\subsection{\it Results and discussion
\label{subsec:disc_results}}

We present in Figs.\ \ref{fig:fus64zn}--\ref{fig:fus11be} fusion
excitation functions calculated using the double-folded potentials
obtained as described above and compare them with data for total
fusion, with the exception of the $^6$He + $^{64}$Zn system where
we plot both the ``raw'' residue cross sections of \cite{DiP04}
which could contain events due to transfer reactions (referred to
as ``total fusion'' for brevity) and the ``corrected'' cross
section where the measured $^{65}$Zn cross section has been
replaced by that calculated with the code \textsc{cascade}
\cite{Puh77} (referred to as ``complete fusion'' for brevity), the
$^6$He + $^{238}$U \cite{Raa04} and $^7$Be + $^{238}$U \cite{Raa06}
systems where we plot the {\em complete} fusion\footnote{Actually
events leading to fission without a charged residue. See also the
discussion in sections \ref{subsubsec:6he238u} and
\ref{subsubsec:7be238u}.} and fission cross sections (the total
fusion cross section should fall somewhere between the two), the
$^4$He + $^{238}$U system where we plot the fission cross sections
\cite{Raa04,Vio62,Zai80} (which should, however, essentially equate
to the total fusion cross section) and the $^6$He + $^{206}$Pb
system where we plot the ($^6$He,$2n$) cross sections of
\cite{Pen06}. In the following discussion enhancement and
suppression of fusion are defined relative to the ``bare'' fusion
calculations denoted by the dotted curves in Figs.\
\ref{fig:fus64zn}--\ref{fig:fus11be}. The calculated nominal
Coulomb barriers are given in Table \ref{tab:coulbars}.
\begin{table}
 \renewcommand{\arraystretch}{1.2}
 \begin{center}
  \caption{Nominal Coulomb barrier heights for the systems
  considered in this section.
  \label{tab:coulbars}}\bigskip
  \begin{tabular}{l c}
   \hline \hline \\
   [-5mm] \ \ System & $V_\mathrm{B}$ \\
   \hline \\
   [-4mm]
   $^{64}$Zn + $^4$He & $9.50$ MeV \\
   $^{64}$Zn + $^6$He & $8.34$ MeV \\
   $^{206}$Pb + $^6$He & $19.04$ MeV \\
   $^{209}$Bi + $^4$He & $21.52$ MeV \\
   $^{209}$Bi + $^6$He & $19.44$ MeV \\
   $^{238}$U + $^4$He & $22.79$ MeV \\
   $^{238}$U + $^6$He & $20.76$ MeV \\
   $^{209}$Bi + $^6$Li & $29.97$ MeV \\
   $^{238}$U + $^7$Be & $43.43$ MeV \\
   $^{209}$Bi + $^{10}$Be & $39.30$ MeV \\
   $^{209}$Bi + $^{11}$Be & $37.12$ MeV \\
   \hline \hline
  \end{tabular}
 \end{center}
\end{table}
\begin{figure}
\epsfysize=9.0cm
\begin{center}
  \includegraphics[width=.65\textwidth]{fus64zn}
\end{center}
\caption{Calculated $^6$He + $^{64}$Zn (left) and $^4$He +
$^{64}$Zn (right) fusion excitation functions compared with the
total fusion data of Di~Pietro {\em et al.\/} \cite{DiP04} (filled
circles), the sum of the $^{64}$Zn($^4$He,$n$),
$^{64}$Zn($^4$He,$p$), $^{64}$Zn($^4$He,$2n$),
$^{64}$Zn($^4$He,$np$) and $^{64}$Zn($^4$He,$n$+$^4$He) data of
Levkovskiy \cite{Lev91} (open circles), and the sum of the
$^{64}$Zn($^4$He,$n$) and $^{64}$Zn($^4$He,$p$) data of Ruddy and
Pate \cite{Rud69} (open triangles) and Porile \cite{Por59} (filled
diamonds). The open circles on the left hand plot denote the
corrected $^6$He + $^{64}$Zn complete fusion data of Di~Pietro {\em
et al.\/} \cite{DiP04} (see text for details). The arrows indicate
the positions of the calculated Coulomb barriers.
\label{fig:fus64zn}}
\end{figure}
\begin{figure}
\epsfysize=9.0cm
\begin{center}
  \includegraphics[width=.5\textwidth]{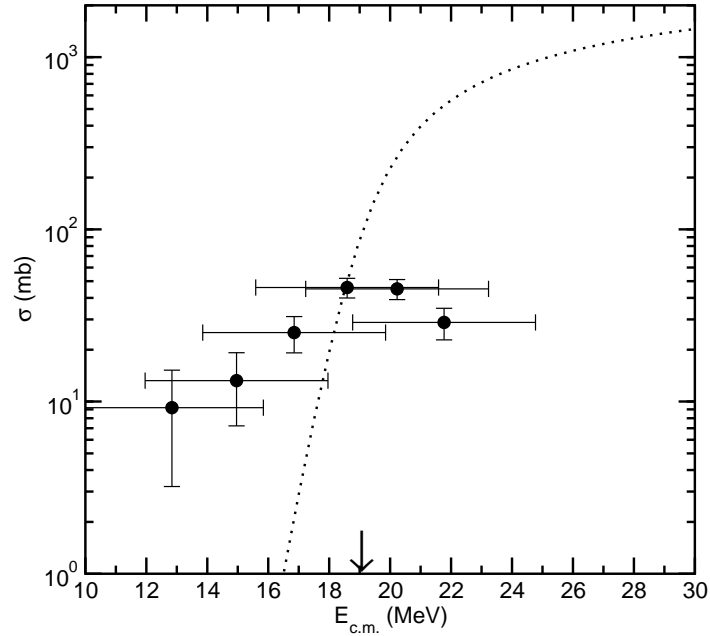}
\end{center}
\caption{Calculated $^6$He + $^{206}$Pb fusion excitation function
compared with the $^{206}$Pb($^6$He,$2n$) data of Penionzhkevich
{\em et al.\/} \cite{Pen06}. The arrow indicates the position of
the calculated Coulomb barrier. \label{fig:fus206pb}}
\end{figure}
\begin{figure}
\epsfysize=9.0cm
\begin{center}
  \includegraphics[width=.65\textwidth]{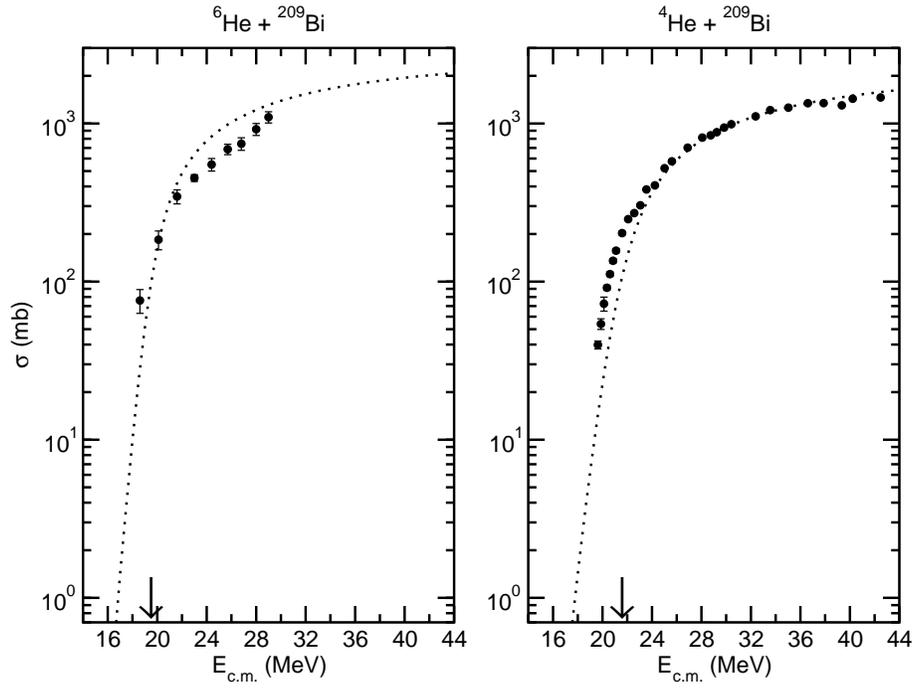}
\end{center}
\caption{Calculated $^6$He + $^{209}$Bi (left) and $^4$He +
$^{209}$Bi (right) fusion excitation functions compared with the
total fusion data of Kolata {\em et al.\/} \cite{Kol98} and
Barnett and Lilley \cite{Bar74} and Ramler {\em et al.\/}
\cite{Ram59}, respectively. The arrows indicate the positions of
the calculated Coulomb barriers. \label{fig:fus209bi}}
\end{figure}
\begin{figure}
\epsfysize=9.0cm
\begin{center}
  \includegraphics[width=.65\textwidth]{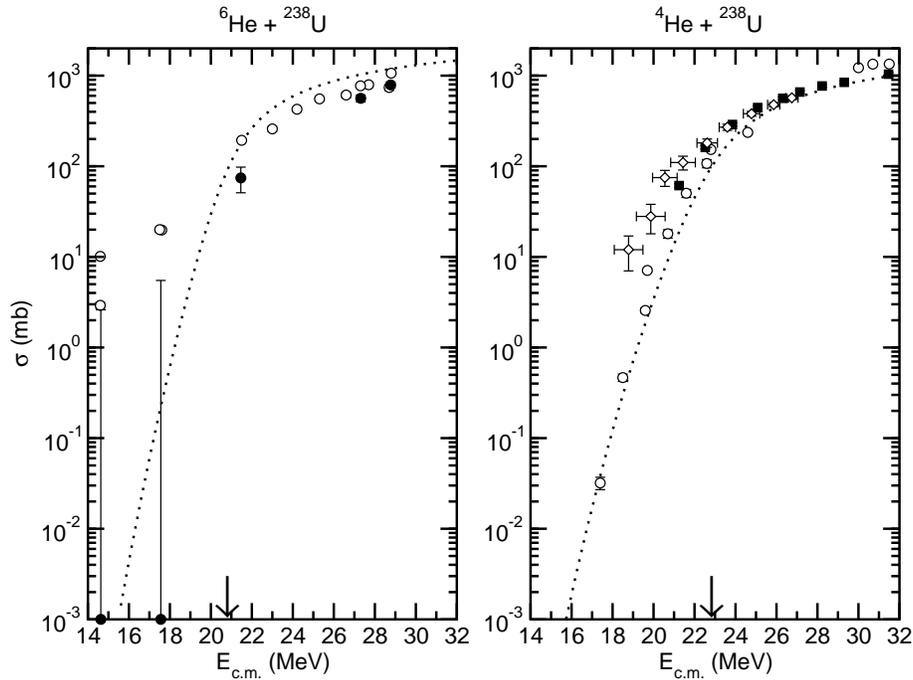}
\end{center}
\caption{Calculated $^6$He + $^{238}$U (left) and $^4$He +
$^{238}$U (right) fusion excitation functions compared with the
{\em complete} fusion (filled circles) and fission (open circles)
data of Raabe {\em et al.\/} \cite{Raa04}. The filled squares and
open diamonds denote the fission data of Viola and Sikkeland
\cite{Vio62} and Zaika {\em et al.\/} \cite{Zai80}, respectively.
The arrows indicate the positions of the calculated Coulomb
barriers. \label{fig:fus238u}}
\end{figure}
\begin{figure}
\epsfysize=9.0cm
\begin{center}
  \includegraphics[width=.65\textwidth]{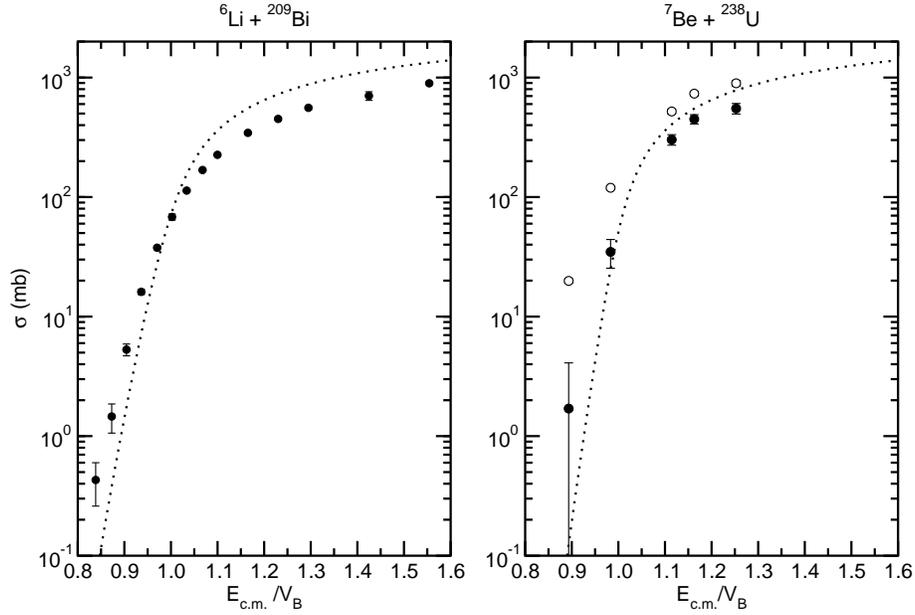}
\end{center}
\caption{Calculated $^6$Li + $^{209}$Bi (left) and $^7$Be +
$^{238}$U (right) fusion excitation functions compared with the
{\em complete} fusion data of Dasgupta {\em et al.\/} \cite{Das04}
and Raabe {\em et al.\/} \cite{Raa06}, respectively. The open
circles denote the fission cross sections of Raabe {\em et al.\/}
\cite{Raa06}. \label{fig:comp6li7be}}
\end{figure}
\begin{figure}
\epsfysize=9.0cm
\begin{center}
  \includegraphics[width=.65\textwidth]{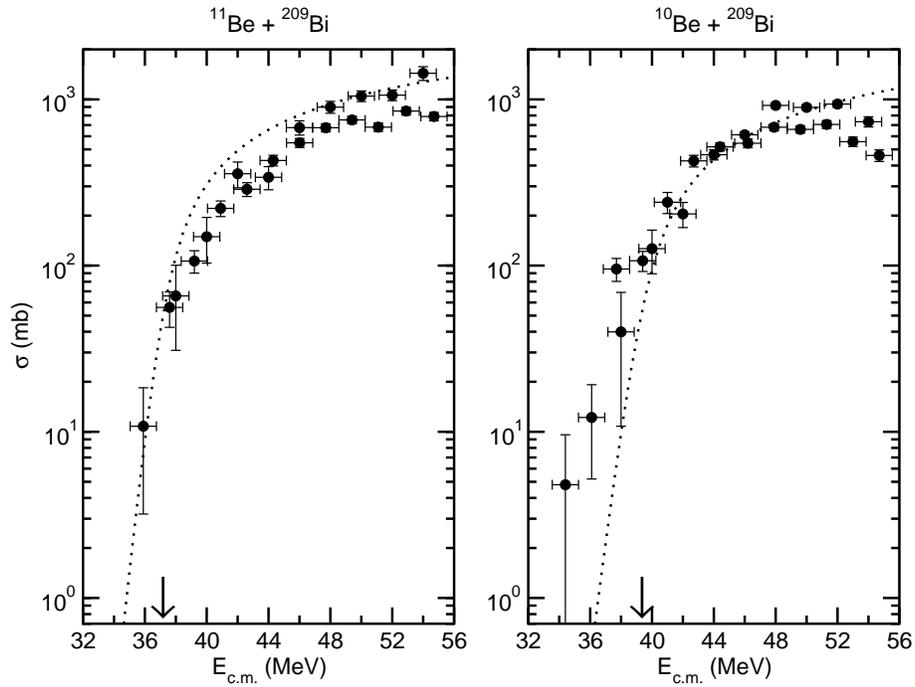}
\end{center}
\caption{Calculated $^{11}$Be + $^{209}$Bi (left) and $^{10}$Be +
$^{209}$Bi (right) fusion excitation functions compared with the
total fusion data of Signorini {\em et al.\/} \cite{Sig98} and
Signorini \cite{Sig02}, respectively. The arrows indicate the
positions of the calculated Coulomb barriers. \label{fig:fus11be}}
\end{figure}

We may draw some general conclusions from these figures, bearing in
mind the caveats concerning the double-folded potentials. With the
possible exception of the $^4$He + $^{64}$Zn system where the
various data sets do not agree very well with each other, the data
for total fusion of the ``core'' nuclei are rather well described
for energies above the Coulomb barrier, taking into account a
certain amount of scatter in the data points for some systems.
Where data points exist for sub-barrier energies the enhancement of
measured to calculated fusion cross sections characteristic of
``normal'' nuclei, see e.g.\ \cite{Bec85,Ste86,Bec88}, is also
observed for these systems, with the possible exception of $^4$He +
$^{238}$U where the different data sets disagree at sub-barrier
energies. The level of agreement obtained for the core fusion
excitation functions at above barrier energies where we expect
coupling effects to be small for these systems, at least for those
with $^4$He projectiles, suggests that our adopted effective
nucleon-nucleon interaction and target matter densities are
physically reasonable.

The total fusion data for the halo nuclei are also consistent with
a possible sub-barrier enhancement of total fusion, although this
conclusion is very tentative, to say the least, as it rests solely
on the lowest energy data points in the $^6$He + $^{64}$Zn, $^6$He
+ $^{206}$Pb and $^6$He + $^{209}$Bi systems. The other systems do
not actually rule out the possibility of sub-barrier enhancement,
either because the data stop short of the energies at which
enhancement would be readily apparent or due to uncertainties in
the data, due to the difficulty of defining unambiguously a fusion
event at these low energies and/or poor statistics due to the low
cross sections. Bearing in mind the uncertainties in the
double-folded potentials discussed above and their effect on the
calculated fusion excitation functions, sub-barrier enhancement of
the total fusion cross section for halo nuclei remains to be
confirmed.

\begin{figure}
\epsfysize=9.0cm
\begin{center}
  \includegraphics[width=.5\textwidth]{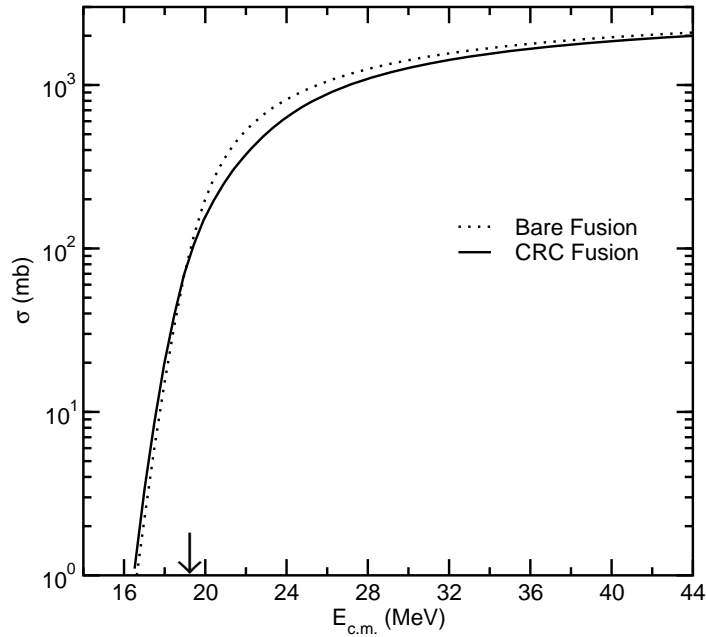}
\end{center}
\caption{Calculated $^{6}$He + $^{208}$Pb total fusion with (full
curve) and without (dotted curve) coupling to the ($^6$He,$^5$He)
stripping. The arrow indicates the position of the calculated
Coulomb barrier. \label{fig:6he208pbe}}
\end{figure}

The most striking general conclusion with regard to total fusion
for the halo nuclei is not, however, concerned with the sub-barrier
regime but rather with the suppression of the measured total fusion
cross sections compared to the ``bare'' no-coupling calculations at
energies a few MeV above the nominal Coulomb barriers. With the
exception of the $^6$He + $^{206}$Pb system where only
$^{206}$Pb($^6$He,$2n$) data are available --- essentially
equivalent to total fusion for energies below the Coulomb barrier
--- and we consequently cannot comment on the above barrier fusion,
all the total fusion excitation functions for the halo systems
studied here exhibit this characteristic behaviour. This is
particularly noticeable for the $^6$He + $^{209}$Bi,$^{238}$U and
$^{11}$Be + $^{209}$Bi systems. This is in complete contrast to the
total fusion excitation functions for the core nuclei, with the
possible exception of the $^4$He + $^{64}$Zn system. This
suppression compared to the ``bare'' calculation for the halo
nuclei is identical to that due to coupling to the single neutron
stripping reaction in the CRC calculations presented in section
\ref{subsec:transfer} for the $^{6,8}$He + $^{208}$Pb systems.
There are hints in the data that the suppression factor peaks
somewhat above the barrier and then diminishes as the bombarding
energy is increased. In Fig.\ \ref{fig:6he208pbe} we present
calculations for the $^6$He + $^{208}$Pb system showing the effect
of coupling to the ($^6$He,$^5$He) stripping over an extended
energy range which exhibit this feature.

It is therefore tempting to conclude that single neutron stripping
reactions are indeed responsible for this important suppression of
the {\em total} fusion of neutron halo nuclei at energies just
above the Coulomb barrier. It should, however, be borne in mind
that our calculations probe the effect of coupling to one reaction
partition --- single neutron stripping --- in isolation. Other
couplings, in particular two neutron transfer and breakup, may
modify the overall effect. Nevertheless, the suggestion that this
large effect due to neutron transfer is a general phenomenon and
not just confined to the fusion of $^{6,8}$He with heavy targets is
borne out by test calculations which show similar effects due to
the ($^6$He,$^5$He) and ($^{11}$Be,$^{10}$Be) couplings in the
$^6$He + $^{58}$Ni and $^{11}$Be + $^{208}$Pb systems. However,
while the transfer couplings do contribute to a sub-barrier
enhancement of the total fusion they are far from being able to
account for the whole of the effect seen in Fig.\
\ref{fig:fus206pb} for the $^6$He + $^{206}$Pb system, for example.
The important influence of transfer couplings on total fusion for
halo nuclei at energies above the barrier is in complete contrast
to the negligible effect these couplings have on fusion for the
stable weakly bound nucleus $^6$Li, see section
\ref{subsec:transfer}. However, it is similar to the effect of
transfer couplings on fusion in the $^{18}$O + $^{58}$Ni system
\cite{Per06}, for example, although there the main above barrier
suppression is due to two neutron transfer.

It is suggestive that fusion induced by $^{18}$O exhibits the same
important influence of transfer couplings on the above barrier
fusion as we appear to see for the halo nuclei and that in both
cases the transfer reactions have positive $Q$-values. However, as
we have seen in section \ref{subsec:transfer} a positive $Q$-value
alone is not sufficient to obtain a significant effect from
coupling to a transfer partition. In fact, the effect of a given
set of couplings on other observables is a complex combination of
$Q$-values, angular momentum matching conditions and spectroscopic
factors, or more generally coupling strengths.

We note that the three-body time-dependent wave-packet model
calculations of \cite{Ito06} show similar above barrier suppression
of the total fusion cross section to that found in our CRC
calculations. However, they see no sub-barrier enhancement; their
results show either essentially no effect or a slight suppression
in the sub-barrier energy regime.

A sub-barrier enhancement due to multiple neutron transfer is
predicted in the Stelson model \cite{Ste90}: according to this
picture, the weak binding of the neutrons in the projectile should
favour the formation of a ``neck'' between the two reacting nuclei,
thus helping to overcome the Coulomb barrier. We should however
underline that this is a phenomenological model, ignoring the
essential coupled-channel nature of the reaction mechanism.

We have so far not discussed the question of {\em complete} fusion
involving halo nuclei and the effect of couplings thereon as with
the exception of the $^6$He + $^{238}$U and $^6$He + $^{206}$Pb
systems the available data are for total fusion. Clearly, the
complete fusion cross sections will be rather smaller in magnitude
than those for total fusion, although exactly how much smaller will
depend on the system under consideration. Thus the suppression at
above barrier energies should be somewhat larger for complete
fusion than for total fusion and the sub-barrier enhancement (if
present) smaller. The additional above barrier suppression for
complete compared to total fusion could be considerable judging by
the data for the $^6$Li + $^{209}$Bi system, cf.\ Figs.\
\ref{fig:fus6li5li} and \ref{fig:comp6li208pb}, although as we have
seen this may not be a reliable guide for the case of halo nuclei.
Other couplings may well be needed to account for any extra
suppression in the complete fusion and an obvious candidate is
breakup. The calculations of \cite{Hag00,Dia02}, for example,
support this conjecture.

In Fig.\ \ref{fig:comp6li7be} we compare the complete fusion cross
sections for the $^6$Li + $^{209}$Bi \cite{Das04} and $^7$Be +
$^{238}$U \cite{Raa06} systems with the calculated bare fusion
excitation functions. Although $^7$Be is weakly bound it is not a
halo nucleus, and is expected to exhibit behaviour similar to
$^6$Li due to its similar breakup threshold \cite{Kee02a}. This
does indeed appear to be the case as far as the complete fusion is
concerned, as we see similar suppression at above barrier energies
to that observed for $^6$Li. However, it is less marked and there
is no evidence of a sub-barrier enhancement similar to that seen in
the $^6$Li + $^{209}$Bi system, although the uncertainties in the
sub-barrier data points for $^7$Be + $^{238}$U do not completely
rule out this possibility. It is not possible to say whether the
above-barrier {\em total} fusion is reasonably well described by
the bare fusion calculation, as in the $^6$Li + $^{209}$Bi case,
although again the available data do not rule out this possibility
(the total fusion cross section should lie somewhere between the
complete fusion and fission cross sections).

We shall now briefly discuss the question of the effect of coupling
to breakup on fusion for halo nuclei. It is clear that any large
sub-barrier enhancement of total fusion compared to a no-coupling
calculation could not be accounted for by coupling to single
neutron stripping, judging by the calculations for the $^6$He +
$^{208}$Pb system presented in Figs.\ \ref{fig:fus6he5he} and
\ref{fig:6he208pbe}. While couplings to inelastic excitation of the
target should contribute to any sub-barrier enhancement of total
fusion, the bulk of any effect may well be due to coupling to
breakup and/or two neutron transfer, although for single neutron
halo systems like $^{11}$Be the latter process is not expected to
be important. Most coupled channels based calculations suggest that
the effect of coupling to breakup on the {\em total} fusion for
sub-barrier energies should be an enhancement, although the extent
of this enhancement and the bombarding energy at which it sets in
depend on the details of the calculation, see e.g.\ \cite{Can06}
for a review of the literature on this subject. The extent of any
enhancement may also depend quite sensitively on the breakup
threshold energy, see e.g.\ \cite{Tak91}. We have already seen in
section \ref{subsec:breakup} that the magnitude of the breakup
cross section and the effect of coupling to breakup on the elastic
scattering angular distribution can be quite sensitive to the
breakup threshold energy, depending on the target nucleus.

The situation for sub-barrier {\em complete} fusion is less clear.
The $^{206}$Pb($^6$He,$2n$) data of \cite{Pen06} might be equated
to complete fusion at sub-barrier energies with a reasonable degree
of confidence, thus indicating a sub-barrier enhancement of the
complete fusion cross section, although this is only really
observed at the lowest energy measured due to the uncertainties in
the incident energies. However, it is less certain whether coupling
to breakup will also lead to a sub-barrier enhancement of complete
fusion. Penionzhkevich {\em et al.\/} \cite{Pen06} accounted for
sub-barrier enhancement in their $^{206}$Pb($^6$He,$2n$) data by a
``sequential fusion'' mechanism \cite{Zag04} with transfer of the
two ``valence'' neutrons of $^6$He followed by fusion of the $^4$He
core. However, the data for events leading to fission without a
charged residue for the $^6$He + $^{238}$U system (probably
equivalent to complete fusion) are consistent with no sub-barrier
enhancement, although the large uncertainties in the sub-barrier
data points --- it is clear that the cross section for these events
is small compared to that for total fission --- mean that the
possibility of enhancement in this system cannot be entirely ruled
out.

However, perhaps the most remarkable feature of reactions involving
halo nuclei is not concerned with fusion at all, but rather direct
reactions. Very large total reaction cross sections at near and
sub-barrier energies have been observed in the $^6$He + $^{209}$Bi
\cite{Agu01} and $^6$He + $^{64}$Zn \cite{DiP04} systems. These
large cross sections have been identified with $\alpha$ production
arising from direct reactions: $^6$He $\rightarrow$ $\alpha+2n$
breakup, ($^6$He,$^5$He) transfer and ($^6$He,$^4$He) transfer.
Indeed, a total $\alpha$ production cross section of nearly 200~mb
was observed for the $^6$He + $^{209}$Bi system at an incident
energy 6~MeV {\em below} the nominal Coulomb barrier \cite{Agu01}.
For the $^6$He + $^{209}$Bi system further exclusive measurements
employing $\alpha+n$ coincidences have further established that the
$\alpha$ production cross section is dominated by transfer;
approximately 75\% \cite{Byc04,DeY05} at an incident energy of
$\sim 23$~MeV with $2n$ transfer contributing the largest share.
The dominant contribution of transfer to the observed sub-barrier
fission yield in the $^6$He + $^{238}$U system \cite{Raa04} is
consistent with this picture. Thus, for halo systems the total
reaction cross section at sub-barrier energies seems to be
dominated by the direct component, fusion giving a negligible
contribution. While the direct reaction cross section is also
observed to be larger than fusion at sub-barrier energies in other
systems, see e.g.\ Fig.\ \ref{fig:fus6li5li} for the $^6$Li +
$^{209}$Bi system and \cite{Vul86} for the $^{16,18}$O + $^{208}$Pb
and $^{15}$N,$^{16}$O + $^{209}$Bi systems, it is the size of the
effect for the halo systems that is remarkable; direct reactions
completely dominate the total reaction cross section at these
energies, unlike in other systems. Test calculations support the
inference that this is a general phenomenon for halo systems;
calculations of the $^{58}$Ni($^6$He,$^5$He)$^{59}$Ni and
$^{208}$Pb($^{11}$Be,$^{10}$Be)$^{209}$Pb reactions exhibit a
similar dominance over the calculated fusion at near and
sub-barrier energies, as do calculations for the
$^{208}$Pb($^8$He,$^7$He)$^{209}$Pb reaction, see Fig.
\ref{fig:fus8he7he}.

In summary, a consistent analysis of the available data suggests
that the total fusion excitation functions for neutron halo nuclei
exhibit common behaviour so far as the limited sample of systems
and range of energies allows a general conclusion to be drawn. The
data are consistent with an above barrier suppression and possibly
a sub-barrier enhancement of the total fusion cross section,
although the latter remains to be confirmed. While the above
barrier suppression may be accounted for in large part by coupling
to single neutron stripping, any sub-barrier enhancement is most
probably due to coupling to two neutron transfer and/or breakup,
depending on the system. However, much more remarkable are the very
large near and sub-barrier direct reaction cross sections,
dominating fusion by orders of magnitude. These also appear to be a
general feature of reactions involving halo systems.

%\newpage

\section{Conclusions
\label{sec:conclusions}}

In this paper we have described the current experimental and
theoretical situation concerning reactions with light unstable
nuclei at energies around the Coulomb barrier. We saw on the one
hand the problems related to the measurements, due to the quality
of the available beams of light radioactive ions, and on the other
hand the difficulties associated with trying to predict the
behaviour of such systems theoretically. The latter are mainly
connected with poor knowledge of the parameters involved in the
description (details of nuclear structure, the positions and
characteristics of excited states, spectroscopic factors, etc.)
and, in the specific case of the coupling of breakup to fusion, to
the lack of an adequate model.

Given the limited sample of systems for which data are currently
available and the large uncertainties in some of the data sets
--- inevitable due to the difficulty of the experiments --- it may
seem premature to attempt to draw general conclusions concerning
reactions involving halo nuclei. Nevertheless, in our
investigations we came across some recurring features. Within the
constraints imposed by the problems mentioned above, we may
summarise these features as follows:
\begin{enumerate}
\item In general, the magnitude of a reaction channel is not a
reliable indicator of the effect that, through coupling, that
channel may have on elastic scattering or fusion. This was shown by
calculations for the case of breakup and elastic scattering. For
fusion, it goes back to the observation made in \cite{Das94} that
the transmitted flux (and thus fusion) and the reflected flux (the
measured exit channels) are determined by different regions of the
radial couplings.

\item Concerning fusion we observe the following: compared to
``bare'' no-coupling one-dimensional barrier penetration model
calculations employing reasonably realistic double-folded
potentials, the measured total fusion data for halo nuclei
consistently show an above barrier suppression. There is some
evidence that the degree of suppression peaks somewhat above the
nominal Coulomb barrier then gradually declines as the incident
energy is increased. The data for total fusion are also consistent
with a possible sub-barrier enhancement, although this remains to
be confirmed. Calculations suggest that the above barrier
suppression may be attributed in large part to the effect of
coupling to single neutron transfer reactions, although breakup
will also contribute \cite{Dia03}. Any significant enhancement at
sub-barrier energies must be due to other couplings, most probably
breakup and/or two neutron transfer (depending on the system).

\item As the {\em complete} fusion cross section will be smaller
than the total fusion we anticipate an increased above barrier
suppression of complete fusion compared to total fusion and a
smaller sub-barrier enhancement, if this is proved to be present.
Depending on the system, couplings in addition to single neutron
transfer may be needed to adequately describe the increased above
barrier suppression.

\item Reactions involving halo nuclei exhibit a remarkable
dominance of direct reactions over fusion at near and sub-barrier
energies. The cross sections for direct reactions can be orders of
magnitude larger than fusion at sub-barrier energies, a phenomenon
that appears to be unique to halo systems. The most important
component of the direct reaction cross section is probably due to
neutron transfers; this has been shown experimentally for the
$^6$He + $^{209}$Bi system and is borne out by calculations for
other systems. Despite the relatively low thresholds, breakup does
not seem to be a very large component of the total reaction cross
section for the halo systems studied here, a conclusion that is
again largely borne out by calculations.
\end{enumerate}

These are our conclusions based on the currently available data for
halo nuclei, but what of the future? Clearly, further data are
highly desirable extending to other halo systems such as $^8$He,
$^{11}$Li, $^{14}$Be, $^{15}$C and others. Post-accelerated beams
of some of these nuclei are already available, for example
$^{11}$Li at TRIUMF/ISAC in Canada. However, the current
intensities ($10^3$--$10^4$ pps) are still too low to allow
measurements of fusion at barrier energies. A significant increase
in intensity will be brought about only by the next generation of
radioactive beam facilities currently under construction or
development. However, much can still be done in the way of
improving our knowledge of nuclei for which beams of higher
intensity (at least $10^5$ pps) are already available. In
particular, measurements of the total fusion for $^6$He at far
sub-barrier energies with good energy definition and reasonable
statistics are needed to decide the question of sub-barrier fusion
enhancement in a definitive manner. Extension of the available data
to systems with lighter mass targets is also desirable, although
the difficulty of defining a reliable signature for fusion in this
mass range could be an obstacle \cite{DiP04,Nav04}. Further use
could be made of available beams of $^8$He --- so far there is a
single measurement of the elastic scattering of $^8$He from a
medium mass target \cite{Nav04} --- which is expected to show an
even larger sub-barrier reaction cross section than $^6$He, cf.\
Figs.\ \ref{fig:fus6he5he} and \ref{fig:fus8he7he}.

Choice of target is also an important consideration from both the
experimental and theoretical points of view. Experimentally, in
measurements of fusion choice of target is often limited by
considerations of a clear signature for fusion. It is unfortunate
that most of the available data for fusion involving halo systems
are thus for targets such that a comprehensive theoretical
analysis including couplings to transfer reactions is not possible,
due to the density of states and lack of the necessary nuclear
structure information. While the experimental constraints are
probably most severe, there should be scope for compromise choices
where a reasonably complete calculation is possible without
compromising the measurement. If the effect of coupling to two
neutron transfer on fusion is to be elucidated, careful choice of
the target will be required.

Further developments in the theory describing fusion of weakly
bound nuclei in general are also required. In particular, more
sophisticated treatments of the incomplete fusion process within a
CDCC-type formalism are needed. The ideal would be to combine such
methods with a CRC calculation of transfer, although we are still
a considerable way from achieving this goal.

To counter the low beam intensities, the various experimental
techniques used for the detection of fusion and direct reactions
have reached in the last few years a very good level of efficiency.
Refinements in the measurement of fusion and direct reactions will
continue with the nuclei currently available at radioactive beam
facilities.

In the future however, further \emph{exclusive} measurements of the
direct reaction processes, such as those already carried out for
the $^6$He + $^{209}$Bi system \cite{Byc04,DeY05}, are needed to
help unravel the competition between breakup and transfer for
neutron halo nuclei. One may eventually hope to have angular
distribution measurements for a more detailed comparison with
calculations. To reach this goal the implementation of different
simultaneous signatures --- charged particles, prompt $\gamma$-ray
emission and neutron detection --- is required. This calls for
collaborations where expertise is brought together in some key
experiments. The choice of the systems to measure will have to take
into account the experimental constraints but also the theoretical
ones as we have illustrated above.

%\newpage

\section*{Acknowledgements
\label{sec:acknow}}

The authors would like to thank Dr.\ A.~Di~Pietro, Prof.\ Yu.\
Penionzhkevich, Prof.\ J.J.~Kolata, Prof.\ C.~Signorini and Dr.\
M.~Dasgupta for providing their data in tabular form; Prof.\
A.~Vitturi and Dr.\ A.~Navin for stimulating and instructive
discussions. N.K. gratefully acknowledges the receipt of a Marie
Curie Intra-European Fellowship from the European Commission,
contract No.\ MEIF-CT-2005-010158. R.R. is a Postdoctoral Fellow of
the Fund for Scientific Research-Flanders (Belgium)
(F.W.O.-Vlaanderen).

%\nocite{*}
\bibliographystyle{prsty}
\bibliography{../references}

\end{document}